\documentclass[useAMS,usenatbib]{mnras}

\usepackage{aas_macros}
\usepackage{amsmath}
\usepackage{amssymb}
\usepackage{booktabs}
\usepackage{graphicx}
\usepackage{epstopdf}
\usepackage{multirow}
\usepackage{multicol}
\usepackage{lscape}

\usepackage{subcaption}
\usepackage[T1]{fontenc}
\usepackage{aecompl}
\usepackage{mathptmx}

\usepackage{hyperref}
\hypersetup{
  colorlinks   = true,  
  urlcolor     = blue, 
  linkcolor    = blue, 
  citecolor    = blue, 
  filecolor    = black  
}

\extrafloats{100}

\def\ha{{H$\alpha$}}

\def\oiii{[\ion{O}{III}]}
\def\nii{[\ion{N}{II}]}

\begin{document}

\title[Kinematics of CO(2-1) emission in nearby AGN]{Nuclear kinematics in nearby AGN.\\ \Large I. An ALMA perspective on the Morphology and Kinematics of the molecular CO(2-1) emission.}

\author[Ramakrishnan et al.]{V.~Ramakrishnan,$^{1}$\thanks{E-mail: vramakrishnan@udec.cl} N.~M.~Nagar,$^{1}$ C.~Finlez,$^{1}$ T.~Storchi-Bergmann,$^{2}$ R.~Slater,$^{1}$
	\newauthor
	A.~Schnorr-M\"uller,$^{2}$ R.~A.~Riffel,$^{3}$ C.~G.~Mundell,$^{4}$ A.~Robinson$^{5}$ \\
	\\
$^{1}$Astronomy Department, Universidad de Concepci\'on, Casilla 160-C, Concepci\'on, Chile\\
$^{2}$Departamento de Astronomia, Instituto de F\'isica, Universidade Federal do Rio Grande do Sul, 91501-970, Porto Alegre RS, Brazil\\
$^{3}$Departamento de F\'isica, Centro de Ci\^encias Naturais e Exatas, Universidade Federal de Santa Maria, 97105-900 Santa Maria, RS, Brazil\\
$^{4}$Department of Physics, University of Bath, Claverton Down, Bath, BA2 7AY, UK\\
$^{5}$School of Physics and Astronomy, Rochester Institute of Technology, 85 Lomb Memorial Dr., Rochester, NY 14623, USA.}

\maketitle

\begin{abstract}
	We present the molecular gas morphology and kinematics of seven nearby Seyfert galaxies obtained from our 230~GHz ALMA observations. The CO J=2-1 kinematics within the inner $\sim30\arcsec$ ($\lesssim9$~kpc) reveals rotation patterns that have been explored using the Bertola rotation model and a modified version of the \textit{Kinemetry} package. The latter algorithm reveals various deviations from pure circular rotation in the inner kiloparsec of all seven galaxies, including kinematic twists, decoupled and counter-rotating cores. A comparison of the global molecular gas and stellar kinematics show overall agreement in the position angle of the major axis and the systemic velocity, but larger discrepancies in the disc inclination. The residual maps obtained with both the methods shows the presence of non-circular motions in most of the galaxies. Despite its importance, a detailed interpretation of the physics responsible for non-circular motions will be discussed in a forthcoming work.
\end{abstract}


\section{INTRODUCTION}

The mechanism that drives the energy output from an active galaxy is understood to be governed by the process of accretion onto a supermassive black hole (SMBH). Several works have shown a correlation between the SMBH mass and the overall properties of the spheroidal component of the host galaxy \citep{Ferrarese2000, Gebhardt2000, Tremaine2002}. One such significant finding is the connection between the black hole mass and the velocity dispersion of the host spheroid \citep{Gultekin2009}, which supports the connection between the build-up of stellar mass of a galaxy and its SMBH. A fundamental problem here is in understanding the processes that govern the evolution of the black hole and its effect on the host galaxy.

Feeding the SMBH requires gas to be driven from galaxy scales down to sub-kpc and finally sub-pc scales: in galaxy discs this effectively requires the gas to lose angular momentum. Mechanisms for driving gas to the nucleus could be secular, e.g. due to the presence of nonaxisymmetric nuclear structures such as the spiral arms with different extent of warping, oval distortions and bars, \citep{Kormendy2004, Maciejewski2004, Hopkins2010}, or due to external influences, e.g. galaxy interactions and major or minor mergers \citep{Sanders1988, Quinn1993, Kauffmann2000, DiMatteo2005, Hopkins2008}. For a review on observations of inflows on different spatial scales see \citet{StorchiBergmann2019}.

An important process by which the SMBH effects the evolution of the galaxy is ``feedback'' -- the deposition of mechanical and/or radiative energy from the SMBH into the interstellar medium. The magnitude of this feedback is regulated in part by the magnitude of the accretion flows onto the black hole. The spatial scale over which this process occurs spans several orders of magnitude in distance. The current understanding of this process in a galaxy from a cosmological perspective remains an enigma. It is, therefore, important to study the effects of the feedback on a galaxy at both small and large scales.

\begin{table*}
	\centering
	\caption{List of sample galaxies. Morphology and the nucleus type are obtained from \citet{Malkan1998}. Source position and the redshift are from SIMBAD \citep[which are in turn obtained from the Two Micron All Sky Survey, hereafter 2MASS;][]{Skrutskie2006}. The luminosity distance and scale was estimated assuming a flat cosmology with H$_0$ = 67.8, $\Omega_M = 0.3$ and $\Omega_\Lambda = 0.7$ \citep{Planck2016}. The position angle (P.A.) and the inclination ($i$) are from the 2MASS catalogue.}
	\label{srcList}
	\begin{tabular}{lccccccccc}
	\toprule
	Source      & Morphology & Seyfert type & R.A.    & Dec.    & $z$ & $D_{\rm L}$ & Scale        & P.A.         & $i$ \\
		    &            &              & (J2000) & (J2000) &     & (Mpc) & (pc~arcsec$^{-1}$) & $(^{\circ})$ & $(^{\circ})$ \\
	\midrule
	NGC~1386          & Sb/c    & Sy2 & 03:36:46.24 & -35:59:57.39 & 0.002905 & 11.4 & 55  &  25 & 69 \\
	NGC~1667          & Sc      & Sy2 & 04:48:37.20 & -06:19:11.47 & 0.015204 & 67.4 & 317 & 165 & 39 \\
	NGC~2110          & Sa      & Sy2 & 05:52:11.40 & -07:27:22.23 & 0.007579 & 35.7 & 170 & 165 & 35 \\
	ESO~428$-$G14$^a$ & S0/a    & Sy2 & 07:16:31.23 & -29:19:29.05 & 0.005554 & 28.0 & 134 & 135 & 53 \\
	NGC~3081          & SB0/a   & Sy2 & 09:59:29.53 & -22:49:34.32 & 0.008040 & 40.6 & 193 &  70 & 47 \\
	NGC~5728$^a$      & SABa    & Sy2 & 14:42:23.93 & -17:15:11.41 & 0.009467 & 45.1 & 214 &  32 & 64 \\
	NGC~7213          & Sa      & Sy1 & 22:09:16.26 & -47:09:59.95 & 0.005869 & 22.8 & 110 &  70 & 20 \\
	\bottomrule
	\multicolumn{10}{l}{$^a$~Morphology and the nucleus type are obtained from \citet{deVaucouleurs1991}.} \\
	\end{tabular}
\end{table*}

There is prolific evidence for molecular gas outflows from luminous active galaxies. This outflowing matter is cold interstellar gas that affects the evolution of the host galaxy by fuelling the star formation and accretion onto the central black hole \citep{Morganti2005, Morganti2016, Alatalo2011, Cicone2014, Tadhunter2014}.

Over the past decades, deep observations of galaxies have provided us with an insight into galaxy evolution and structure formation in the Universe. Following these results, the enhanced star formation and nuclear activity in galaxies have been revealed in greater detail by high-resolution broadband imaging and long-slit spectroscopic surveys. With technological advancements, we are now in the era of integral-field spectroscopy, which provides a high-resolution both in the spectral and spatial domain. It is thus possible to efficiently map the ionised gas, stars, star formation rates, and other parameters with a high precision using integral-field units (IFU). In order to improve our understanding of the physical processes and dynamical structures in galaxies, it is critically necessary to combine the stellar and ionised gas kinematics obtained with IFUs with those of molecular gas; the imaging spectroscopy of millimetre (mm) interferometers, such as the Atacama Large Millimeter Array (ALMA) provides the latter.

Several previous works have studied the kinematics of ionised and/or molecular gas in nearby galaxy samples. The best example for the former are the SAURON \citep{deZeeuw2002} and CALIFA surveys \citep{Sanchez2012} that exploited the IFU on the William Herschel and Calar Alto telescopes, respectively. The kinematics of the molecular (typically CO) emission was studied in detail by surveys like NUGA \citep{GarciaBurillo2003}, ATLAS$^{\rm 3D}$ \citep{Cappellari2011, Young2011, Alatalo2013} and EDGE-CALIFA \citep{Bolatto2017, Levy2018} using IRAM and CARMA observations. A majority of the galaxies show a disc-like morphology in which rotation dominates, though non-axisymmetric kinematics are often seen, and sometimes even dominate. Although these previous surveys cover a large number of galaxies, the quality of their kinematic results were hampered by the limited sensitivity and mapping capability of the interferometers used. The study of feedback in galaxies got a significant boost with the comissioning of the ALMA interferometer. The superior CO maps from ALMA in Seyfert galaxies like NGC~1566 revealed noncircular motions from a bar, indicating strong negative torques on gas at $r = 100-300$~pc \citep{Combes2014, Slater2018}. High velocity outflows have also been observed in other galaxies like NGC~1433 \citep{Combes2013}, NGC~1377 \citep{Aalto2012} and IC~5063 \citep{Morganti2015}.

This paper is organised as follows. We discuss the overall objectives of our project and the sample selection in Section~2, and the observations and the data processing in Section~3. The models utilised in this work are discussed in detail in Section~4 followed by the results in Section~5. The discussion and the summary of this work are given in Section~6 and detailed descriptions of the results for each sample galaxy are listed in Section~7.

\section{OBJECTIVES AND SAMPLE SELECTION}

The growing evidence for a correlation between nuclear dust structures and accretion activity supports the hypothesis that nuclear spirals are a mechanism for fuelling the SMBH \citep[e.g.][]{Emsellem2001, Maciejewski2002, Crenshaw2003, Fathi2005, SimoesLopes2007}. Our group has analysed optical and infrared IFU data of several active galaxies, selected for having dusty nuclear structures. These observations reveal streaming motions of ionised and molecular gas towards the nucleus along dusty spiral arms and/or ionised gas outflows \citep[e.g.][]{Riffel2008, Riffel2013b, SchnorrMuller2014a, Diniz2015, SchnorrMuller2017a, Humire2018}.

We thus undertook a pilot study of nine active galaxies that were selected for: (i) showing signatures of strong streaming inflows, outflows, and/or bright (IR) molecular gas discs in our previous IFU observations; (ii) being close enough to resolve structures at scales of tens of parsecs; and (iii) (typically) having previous detections of nuclear CO emission at a lower resolution. These galaxies were observed using ALMA to trace the CO J=2-1 emission.

Thus, in conjunction with our IFU maps of the stellar and ionised gas morphology and kinematics, and high-resolution NIR imaging, we attempt to quantify the high-resolution CO(2-1) observations of seven nuclei from our sample in this work. The remaining two, viz. NGC~1566 \citep{Slater2018} and NGC~3393 \citep{Finlez2018}, have already been analyzed in detail. The sample galaxies, and their basic parameters, are listed in Table~\ref{srcList}.

\begin{table*}
	\centering
	\caption{Observing Parameters.}
	\label{obsPar}
	\begin{tabular}{lccccccc}
	\toprule
	Source      & Date of observation & Flux calibrator & Bandpass / Phase calibrator & Total observing time & Cell size & FWHM$^{a}$                 & rms$^{b}$   \\
		    & 			  &                 &                             & (min)                & (\arcsec) & (\arcsec $\times$ \arcsec) ($^{\circ}$) & (mJy~km~s$^{-1}$) \\
	\midrule
	NGC~1386      & 2014 May 27 & Neptune    & J2357-5311 / J0334-4008 & 39.0 & 0.08 & 0.80 $\times$ 0.54 (85)  & 0.8 \\
	NGC~1667      & 2014 Jul 08 & J0423-013  & J0423-0120              & 45.4 & 0.09 & 0.55 $\times$ 0.51 (66)  & 0.7 \\
	NGC~2110      & 2015 May 14 & Ganymede   & J0423-0120 / J0541-0541 & 45.0 & 0.06 & 0.99 $\times$ 0.56 (-69) & 0.9 \\
	ESO~428$-$G14 & 2016 May 12 & J0750+1231 & J0538-4405 / J0648-3044 & 35.7 & 0.06 & 0.80 $\times$ 0.66 (-57) & 0.8 \\
	NGC~3081      & 2016 May 02 & J1107-4449 & J1037-2934 / J0927-2034 & 36.2 & 0.06 & 0.82 $\times$ 0.59 (-62) & 1.5 \\
	NGC~5728      & 2016 May 15 & Titan      & J1517-2422 / J1448-1620 & 34.4 & 0.06 & 0.65 $\times$ 0.56 (-69) & 0.7 \\
	NGC~7213      & 2014 May 27 & Neptune    & J2056-4714 / J2235-4835 & 44.0 & 0.06 & 0.62 $\times$ 0.57 (81)  & 0.8 \\
	\bottomrule
	\multicolumn{8}{l}{$^{a}$ Major and minor axes of the synthesized beam, and its Position Angle (North to East).} \\
	\multicolumn{8}{l}{$^{b}$ Noise per 2.6~km~s$^{-1}$ channel.} \\
	\end{tabular}
\end{table*}

\section{OBSERVATIONS AND DATA REDUCTION}

All galaxies were observed with ALMA during Cycles 2 (project-ID 2012.1.00474.S; PI - Nagar) and 4 (project-ID 2015.1.00086.S; PI - Nagar). The observations were taken using the ALMA Band 6 receivers on thirty-two 12-meter antennas. Four spectral windows (SPWs), two in the lower sideband and two in the upper sideband were used. Three of these were configured to cover the following lines at relatively high resolution ($\sim 2.6$~km~s$^{-1}$): $^{12}$CO(2-1) ($\nu_{\rm obs}$ = 229.401922 GHz) , $^{13}$CH3OH ($\nu_{\rm obs}$ = 241.548041 GHz) and CS J=5-4 ($\nu_{\rm obs}$ = 243.728532 GHz). The remaining SPW was set to lower spectral resolution in order to better detect continuum emission. The SPWs were thus centred on 229.415 GHz, 227.060 GHz, 241.554 GHz and 243.735 GHz, with bandwidths of 1.875~GHz, 2.0~GHz, 1.875~GHz and 1.875~GHz, respectively, and spectral resolutions of 2.6~km~s$^{-1}$, 20.53~km~s$^{-1}$, 2.6~km~s$^{-1}$, and 2.6~km~s$^{-1}$, respectively.

The flux and the phase calibrators used for every source along with the total observing time and the synthesised beam size are listed in Table~\ref{obsPar}. Data were calibrated and imaged using CASA 4.7.0 \citep{McMullin2007}. For every source, the continuum emission in the uv-plane was subtracted from the spectral windows using the task uvcontsub. The cell sizes used for every source while imaging is given in Table~\ref{obsPar}. The CO(2-1) emission line was strongly detected in all sources over a velocity range of $\sim \pm200$ km s$^{-1}$, and we were able to map the CO line at a channel spacing of 2.6~km~s$^{-1}$.

Strong nuclear continuum emission was detected in all sources except NGC~1667. Therefore, it was necessary to subtract the continuum emission from the $uv$ data before imaging the molecular line emission. To improve the fidelity and dynamic range of the final data cube, both amplitude and phase self-calibration solutions that were obtained during the process of making the continuum maps was applied to the spectral line data before imaging. The CO(2-1) spectral line was then imaged using a natural weighting at a spectral channel spacing of $2.6~{\rm km~s^{-1}}$ for all sources.

\subsection{Moment maps}
\label{momMaps}

The data cubes obtained above are a series of images stacked along a spectral dimension. In order to facilitate further interpretation, it is useful to create 2D maps through a linear combination of individual planes. In this work, we generate three such maps (as is a common practice in the field of feedback studies) with each dependent on the sum (integrated intensity, $M_0$), average (integrated velocity, $M_1$) and the standard deviation (velocity dispersion, $M_2$) along the frequency dimension. The respective maps and their errors were generated using eq.~\ref{momEqs}. The maps are computed with a threshold that defines the pixel values to be included. For all the sources in this work, we generated moment maps assuming a threshold of $4\sigma$ with the noise level shown in Table~\ref{obsPar}. All channels with visible CO emission (typically 250 to 600 channels) were used in the calculation of the moment maps. We note that the galaxies discussed in this work primarily show rotation dominated, rather than dispersion dominated, kinematics. Thus
the standard deviation ($M_2$) maps trace both the intrinsic dispersion of the gas, and the rotation velocity gradient across the synthesized beam. For simplicity, we refer to the $M_2$ maps as the `velocity dispersion' maps, even though they show the upper limit of the intrinsic velocity dispersion especially in the nuclear regions where the rotation velocity gradients are relatively large.

In equation~\ref{momEqs}, the intensity at a given pixel for a channel velocity is given by $I(\nu)$ and the channel separation (or velocity width) by $\Delta\nu$. The term $w$ correponds to $\sum I(\nu)$.

The moment maps of NGC~1386 are shown in Figure~\ref{NGC1386pack} while those of other sources are shown in Appendix~\ref{ALMAmaps}. For all sources except NGC 1386, the integrated CO map peak(s) are not cospatial with the 230~GHz continuum peak (assumed to be the nuclear position). Sources NGC~1667 (Figure~\ref{NGC1667pack}) and NGC~5728 (Figure~\ref{NGC5728pack}) contain two such CO-bright regions within the inner $\sim2\arcsec$\ which symmetrically straddle the nucleus. Spiral arms are evident in the integrated CO emission maps of all sources except for ESO~428-G14 and NGC~5728. These arms, shown as green contours in the Hubble Space Telescope (HST) images, seem to overlap quite well with the spiral dust lanes seen in the HST images.

The moment 1 maps of all sources show a clear velocity gradient with an indication of a rotating component. This motion coupled with the disc-like rotation that can immediately be inferred from the maps of ESO~428-G14 and NGC~5728 gives rise to the classic double-horn shape of the integrated spectrum (Figures~\ref{ESO428pack} and \ref{NGC7213pack}).

\begin{equation}
\begin{gathered}
	M_{0} = \Delta \nu \sum I(\nu) \\
	M_{0, \rm err} = \sqrt{\sum I_{\rm err}^2(\nu)} \\
	M_{1} = \frac{\sum \nu I(\nu)}{w} \\
	M_{1, \rm err} = \sqrt{\sum \bigg(\frac{w \nu - \sum \nu I(\nu)}{w^2}\bigg)^2 \times I_{\rm err}^2(\nu)} \\
	M_{2} = \sqrt{ \frac{\sum (\nu - M_{1})^2 I(\nu)}{w} } \\
	M_{2, \rm err} = \bigg\{\bigg(\frac{w(\nu - M_1)^2 - \sum(I(\nu)(\nu - M_1)^2)}{w^2}\bigg)^2 \times I_{\rm err}^2(\nu) \\
	+ \bigg(\frac{2\sum(I(\nu)(\nu - M_1))}{w}\bigg)^2 \times M_{1,\rm err}^2 \bigg\}^{\!1/2} \\
\end{gathered}
\label{momEqs}
\end{equation}

\subsection{230~GHz Continuum Maps}

The peak flux of the 230~GHz continuum emission of the sample galaxies is in the range $\sim1-5$~mJy. NGC~1386, NGC~5728 and NGC~7213 show an unresolved 230~GHz core, while ESO~428-G14, NGC~2110 and NGC~3081 show extended structure. The elongation of this continuum structure is seen almost along the major axis of the disc in the case of ESO~428-G14 \citep{Falcke1996} and NGC~2110 \citep{Mundell2000}; in the latter case we are clearly detecting the well known S-shaped jet previously seen at lower frequencies. In NGC~3081 the continuum is extended along the minor axis culminating in a secondary shock feature \citep{Nagar1999b}. The two shocked regions thus visible in the continuum map of NGC~3081 could correspond to the two brightest features seen in the nuclear region of the HST map. The only source with no detection of 230~GHz continuum emission is NGC~1667.

\begin{figure*}
	\centering
	\begin{subfigure}{0.33\textwidth}
	\includegraphics[width=0.85\linewidth]{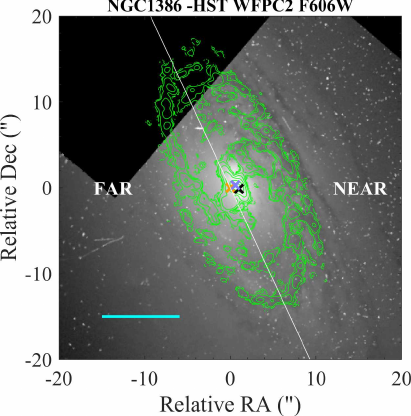}
	\end{subfigure}
	\begin{subfigure}{0.33\textwidth}
	\includegraphics[width=\linewidth]{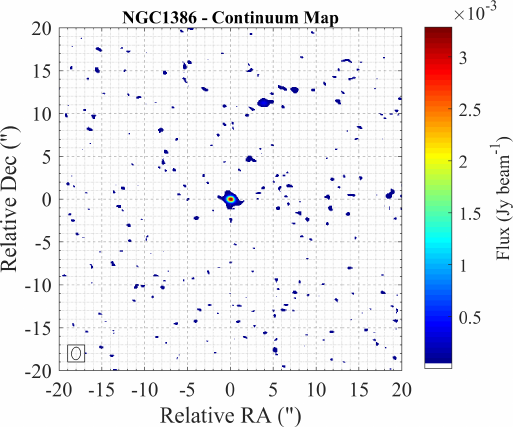}
	\end{subfigure}
	\begin{subfigure}{0.33\textwidth}
	\includegraphics[width=0.85\linewidth]{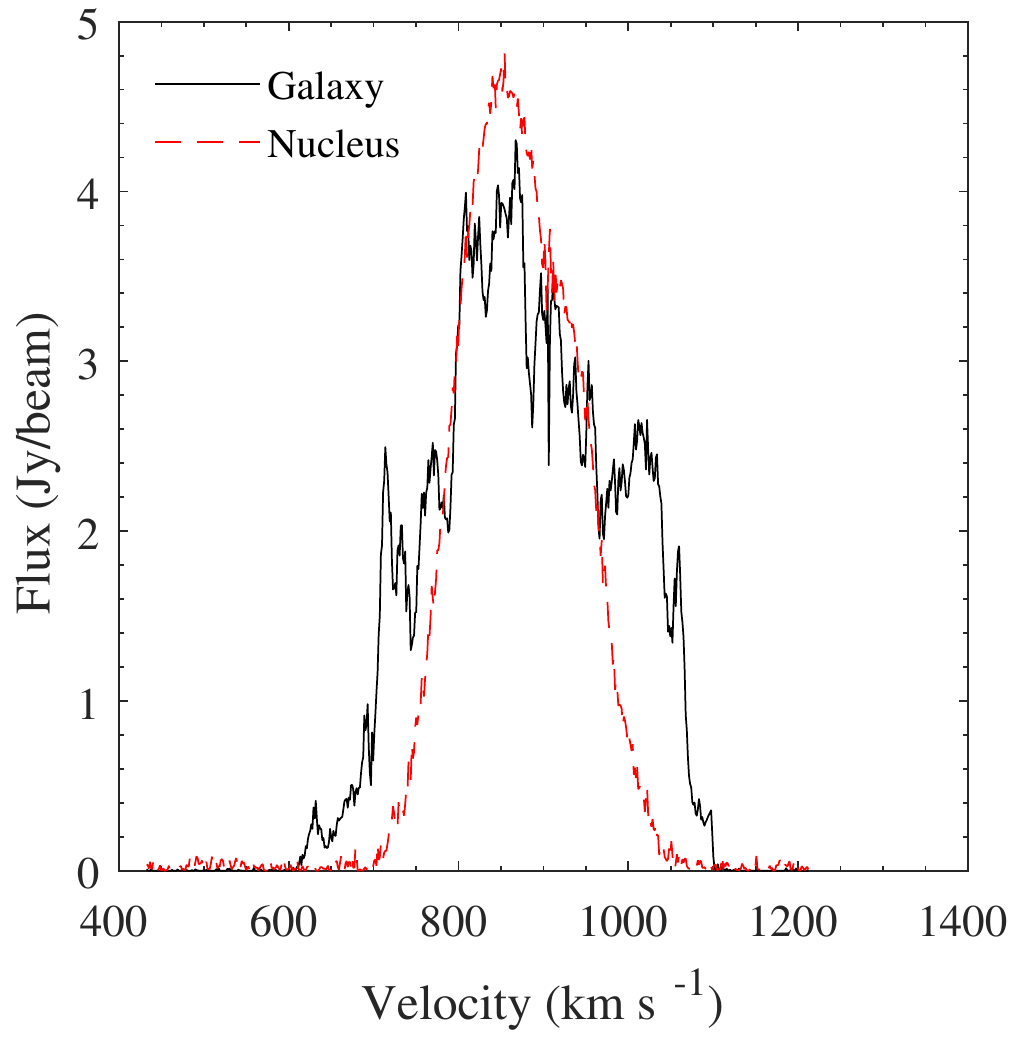}
	\end{subfigure}
	\begin{subfigure}{0.33\textwidth}
	\includegraphics[width=\linewidth]{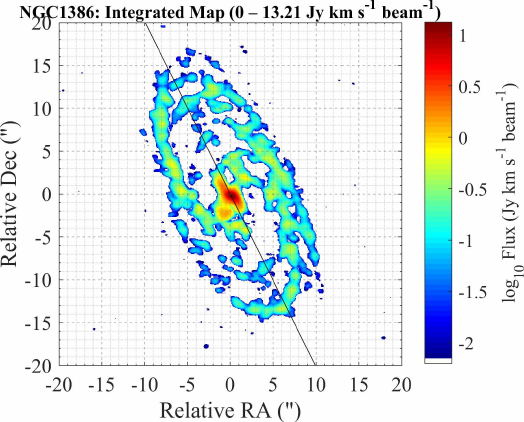}
	\end{subfigure}
	\begin{subfigure}{0.33\textwidth}
	\includegraphics[width=\linewidth]{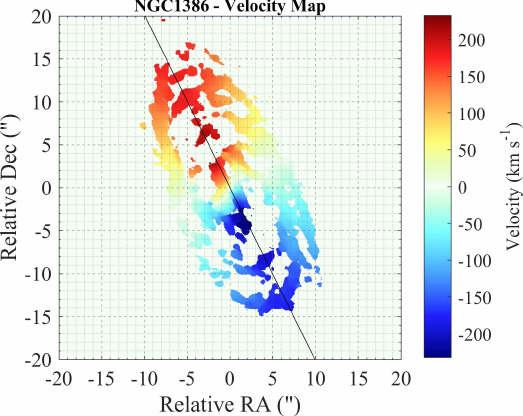}
	\end{subfigure}
	\begin{subfigure}{0.33\textwidth}
	\includegraphics[width=\linewidth]{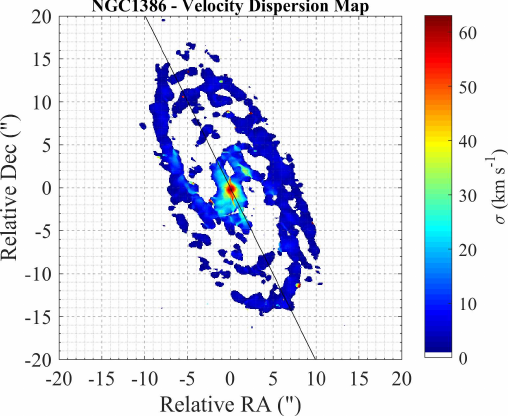}
	\end{subfigure}
	\caption{NGC~1386. \textit{Top-row, from left to right:} A HST image (filter provided in the title), ALMA 230~GHz continuum map, and the galaxy-integrated CO(2-1) spectrum (black) along with the nuclear spectrum (red) extracted from an aperture of radius equal to half the size of the major axis of the synthesised beam. \textit{Bottom-row:} Maps of the integrated flux in logarithmic scale (Jy~km~s${^{-1}}$), velocity (km~s${^{-1}}$) and velocity dispersion (km~s${^{-1}}$) of the CO(2-1) line. The crosses in the HST map corresponds to the centres from the 230~GHz continuum (purple), CO(2-1) kinematics (red), HST optical peak (black) and SIMBAD-listed position (orange), while the green contours overlayed is the respective CO(2-1) integrated map. The white line in the HST map is the photometric major axis while the black one in the moment maps (bottom-row) corresponds to the kinematic major axis. The blue horizontal solid line in the bottom-left corner of the HST image is a scale bar that corresponds to 500~pc.}
	\label{NGC1386pack}
\end{figure*}

\section{Kinematic decomposition methods}

\subsection{Bertola model}

We first model the velocity maps of all galaxies with a pure-rotation model. Specifically, we assume circular orbits in a spherical potential as given in \citet{Bertola1991}.
\begin{equation}
\begin{gathered}
	V_{\rm mod}(R,\psi) = V_{\rm sys} + \\
	\frac{A R \cos(\psi - \psi_0) \sin(\theta) \cos^p(\theta)}{\{R^2 [\sin^2(\psi - \psi_0) + \cos^2\theta \cos^2(\psi - \psi_0)] + c_0^2 \cos^2\theta\}^{p/2}}
	\label{BertolaEq}
\end{gathered}
\end{equation}
Here, $V_{\rm sys}$ is the systemic velocity, $A$ the amplitude of the rotation curve, $R$ and $\psi$ are the radial and angular coordinates for a given pixel in the plane of the sky, $\psi_0$ is the position angle of the line of nodes and $\theta$ is the inclination of the disc. The parameter $p$ denotes the mass distribution of the galaxy such that for a $p = 1$ the rotation curve is asymptotically flat while enclosing the total mass of the system for a $p = 1.5$. The concentration parameter $c_0$ corresponds to the radius at which the velocity reaches 70\% of the maximum amplitude.

In order to obtain a useful interpretation of the gas velocity field, we fit equation~\ref{BertolaEq} to the velocity map as obtained in Section~\ref{momMaps}. The model fitting was done using an optimisation routine that is discussed in the following section (\ref{mads}) allowing five of the six parameters of the model to vary freely, with the sixth ($p$) allowed to vary in the range $1.0 - 1.5$. Apart from the six parameters listed explicitly in equation~\ref{BertolaEq}, we also attempted to find the best-fitting centre coordinate ($x_0, y_0$) of the model. The best-fitting parameters were obtained by minimising a root-mean-square value of the residuals and are reported in Table~\ref{kinResults}. The interpretation of the results of this Bertola fit will be discussed in Section~\ref{kinCent}.

\subsection{Kinemetry}

The velocity model discussed above yields a best fit to the global rotation kinematics of a galaxy. It is also vital to quantify the changes in the kinematic features with radius, using methods which model the kinematics in individual rings \citep[e.g.][]{Begeman1987, Jedrzejewski1987}. Such a method was generalised and established using a technique of Fourier harmonic decomposition in the software \textit{Kinemetry} \citep{Krajnovic2006}. The method, in general, models the moment maps (velocity or higher-order moments) as a series of cosine and sine terms depending on the symmetry (even) or antisymmetry (odd) of the moments. The formulation of this method is given as:
\begin{equation}
	V(a, \psi) = A_0(a) + \sum_{n = 1}^{N} A_n(a) \sin(n\psi) + B_n(a) \cos(n\psi),
	\label{kinEq}
\end{equation}
or in a more compact form as:
\begin{equation}
	V(a, \psi) = A_0(a) + \sum_{n = 1}^{N} k_n(a) cos[n(\psi - \phi_n(a))],
\end{equation}
where the length of the semi-major axis of the ellipse is denoted by $a$ and the eccentric anomaly by $\psi$. The amplitude and phase coefficients, $k_n$ and $\phi_n$, can be estimated from the following relation:
\begin{equation}
	k_n = \sqrt{A_n^2 + B_n^2}; \quad \phi_n = \arctan\bigg(\frac{A_n}{B_n}\bigg).
\end{equation}
In addition to the kinematic coefficients, the code also gives the position angle ($\Gamma$) of the kinematic major axis and the flattening parameter ($q$; ratio of semi-minor to semi-major axes of an ellipse) along the best-fitting elliptical rings. The mode of operation of the algorithm for every ellipse (given either as the number of rings or ring spacing) proceeds in a nested fashion by first extracting the kinematic profile for the first four terms ($A_1, B_1, A_2, B_2$ in addition to $A_0$) from the Fourier expansion given in Equation~\ref{kinEq} to yield a precise estimate of the position angle and the flattening. This is due to the sensitivity of the first four coefficients to the centre coordinates, position angle and the ellipticity ($\epsilon = 1 - q$). For a visual understanding of this concept, please refer to Figure~1 in \citet{Ciambur2015}. Upon obtaining the best-fitting ($\Gamma, q$) pair, the decomposition is repeated for the same ellipse to obtain the values of the required number of Fourier coefficients.

\subsection{Modifications to Kinemetry}

Here, we elucidate the modifications that we have made in the original version of the algorithm discussed above.
\begin{itemize}
	\item The version introduced by \citet{Krajnovic2006} was written in the IDL programming language. We instead adopted the algorithm into \textsc{MATLAB} environment just for the convenience of the authors.
	\item One of the optional input parameters \textit{cover} in the original version corresponds to the threshold for the ratio of the number of pixels with velocity information to the total number of pixels along a sampled ellipse. On reaching this threshold level the iteration stops without moving to the ellipse(s) placed further. Since there are cases with the availability of information beyond a certain radius, i.e. systems with multiple rings consisting of voids in between, it would be desirable to continue the iteration to the next loop instead of terminating it.
	\item A major alteration to the original code is with regard to the eccentric anomaly. This factor corresponds to the shape of the ellipse being sampled. The original code attempts to divide an ellipse into equal bins in angle (i.e. $\phi \in [0; 2\pi]$; where $\phi$ is the azimuthal angular coordinate). While this binning works well for a circle, an ellipse is relatively coarsely sampled along the semi-major axis. This effect can be negligible for many galaxies but gets more pronounced in highly inclined systems. To circumvent this problem, \citet{Ciambur2015} introduced the following transformation:
		\begin{equation}
			\psi = -\arctan\bigg(\frac{\tan(\phi)}{1 - \epsilon}\bigg).
		\end{equation}
		For details on why this formulation creates significant improvements, please see the work of \citet{Ciambur2015}.
	\item In the original code the best-fitting parameters of the position angle and flattening were obtained by first finding the lowest value of the objective function from various pairs of ($\Gamma, q$). The pair corresponding to the lowest value is fed to a Levenberg-Marquardt optimisation routine to find a possible global minimum. We have revised this approach into a single step process by skipping the initial step of surveying using a grid. Instead we focus on obtaining the global minimum for a given bound constraint for the position angle ($\Gamma \in [-90, 90]$) and the flattening ($q \in [0, 1]$). The optimisation is performed using a variant of the direct search routine that is more robust to yield global results. For more details regarding the algorithm see the section below.
\end{itemize}

We refer to our modified version of the Kinemetry as \textit{modKin} in the following sections. The velocity map and its associated error map ($M_1$ and $M_{\rm 1, err}$) were used to quantify the local kinematic features in every galaxy. The velocity map was inversely weighted (1/$(M_{\rm 1, err})^2$) while solving the system of linear equations. Apart from the maps, we provided the \textit{COVER} parameter of 0.50 as input along with the boundary values for the centre coordinates. The number of Fourier coefficients required from the fit was set to four to obtain the best-fitting position angle, flattening and the centre coordinates. Upon obtaining these values, we ran the \textit{modKin} code again at these fixed values for ten Fourier terms. In all cases, the number of ellipses to be fit was initially set to 100 with a ring spacing set according to the maximum size of the interferometric beam.

\subsubsection{Optimisation algorithm}
\label{mads}

The two different models (Bertola and Kinemetry) that are discussed above rely heavily on the use of an optimisation algorithm to obtain the best-fit values. Since both models are non-linear, it is essential to implement a routine that effectively explores the parameter space to converge to a global minimum. We, therefore, adopt a variant of the direct search method, which is a derivative-free optimisation strategy that is becoming quite indispensable in practice \citep{Audet2017}.

The mesh adaptive direct search (MADS) is one such method that generates meshes of the parameter space \citep{Audet2006}. It then performs an adaptive search on the generated trial points as per the tuning parameters, which controls the effectiveness of the exploration, to minimise the given objective function. The nonlinear optimisation coupled with the MADS algorithm guarantees a much faster convergence on the global minimum \citep{Digabel2011}. We therefore use the MADS\footnote{https://www.gerad.ca/nomad/Project/Home.html} algorithm for all our analysis.

The choice of the dimensionality is one of the primary concerns in choosing any optimisation routine. We tested the adopted optimisation algorithm with the Bertola model and \textit{modKin} code on 1000 different simulated velocity maps with known sets of parameters. In all our simulations, the optimisation routine was able to recover the parameters to within an error of 1\%. We, however, caution the reader that this effectiveness is guaranteed only for the lower dimensionality, i.e. we have a maximum of eight parameters to fit in a given model in this work. For models with more parameters, rigorous analysis with the optimisation routine will be required.

\subsection{3D-modelling algorithms}

Algorithms for modelling the spectral cube directly exists in literature. These packages have the advantage of taking into the account the systematics from the observations effectively. Some of the well-known packages are GalPaK \citep{Bouche2015}, $^{3{D}}$BAROLO \citep{DiTeodoro2015} and TiRiFiC \citep{Jozsa2007}. In the ALMA and IFU era where the resolution plays a major role it is essential to start implementing these 3D approaches over the 2D ones. However, for this work, we only consider the 2D modelling approach owing to the lack of features in the mentioned 3D packages to constrain or demarcate the non-axisymmetric features effectively.

\section{RESULTS}

\begin{table*}
	\centering
	\caption{Results from the kinematic analysis.}
	\label{kinResults}
	\begin{tabular}{lccccccccccccc}
	\toprule
	Source & \multicolumn{3}{c}{P.A.$_{\rm kin}$}  & \multicolumn{3}{c}{\textit{i}$_{\rm kin}$} & \multicolumn{3}{c}{$V_{\rm sys}$}  & $\Psi$ & $A$ & $c_0$ & $p$ \\
		& \multicolumn{3}{c}{$(^{\circ})$}  & \multicolumn{3}{c}{$(^{\circ})$} & \multicolumn{3}{c}{(km~s$^{-1}$)}  & ($^{\circ}$) & (km~s$^{-1}$) & (arcsec) &  \\ 
	\cmidrule(lr){2-4} \cmidrule(lr){5-7} \cmidrule(lr){8-10} \cmidrule(lr){12-14}
		& (1) & (2) & (3) & (1) & (2) & (3) & (1) & (2) & (3) &  &  &  & \\
	\midrule
	NGC~1386      & 88$^{b}$  & 26  & 25  & 65 & 69 & 67 & 810  & 884  & 876  &  1.41 & 397 & 1.96 & 1.30 \\
	NGC~1667      & 165$^{c}$ & 160 & 160 & 48 & 56 & 43 & 4570 & 4487 & 4479 &  4.03 & 623 & 1.67 & 1.53 \\
	NGC~2110      & 171$^{d}$ & 171 & 178 & 39 & 51 & 63 & 2309 & 2352 & 2335 &  6.67 & 512 & 1.38 & 1.36 \\
	ESO~428$-$G14 & 119$^{a}$ & 119 & 121 & 54 & 49 & 50 & 1752 & 1669 & 1672 & 15.23 & 376 & 4.04 & 1.33 \\
	NGC~3081      & 90$^{e}$  & 260 & 269 & 40 & 41 & 58 & 2394 & 2385 & 2362 & 10.51 & 384 & 2.44 & 1.43 \\
	NGC~5728      & 30$^{f}$  & 191 & 192 & 55 & 39 & 49 & 2836 & 2763 & 2763 & 20.66 & 419 & 1.17 & 1.25 \\
	NGC~7213      & 305$^{g}$ & --  & 351 & 25 & -- & 42 & 1648 & --   & 1783 &    70 & 174 & 0.31 & 0.95 \\
	\bottomrule
	\multicolumn{14}{l}{\textbf{Notes:} P.A.$_{\rm kin}$ -- position angle; $i_{\rm kin}$ -- inclination; $V_{\rm sys}$ -- systemic velocity. Columns (1) -- (3): stellar kinematics from literature, median values from} \\
	\multicolumn{14}{l}{\textit{modKin} and Bertola. Misalignment angle ($\Psi$) obtained from the photometric P.A. (reported in Table~\ref{srcList}) and the kinematic one} \\
	\multicolumn{14}{l}{is also shown along with other parameters ($A, c_0$ and $p$) from the Bertola model. The uncertainties of the reported values are within 5\%.} \\
	\multicolumn{14}{l}{References: $^{a}$ \citet{Riffel2006}; $^{b}$ \citet{Lena2015}; $^{c}$ \citet{SchnorrMuller2017a}; $^{d}$ \citet{SchnorrMuller2014a};} \\
	\multicolumn{14}{l}{$^{e}$ \citet{SchnorrMuller2016}; $^{f}$ \citet{Emsellem2001}; $^{g}$ \citet{SchnorrMuller2014b}} \\
	\end{tabular}
\end{table*}

\subsection{Kinematic centres}
\label{kinCent}

\begin{figure*}
	\centering
	\begin{subfigure}{0.45\textwidth}
	\includegraphics[width=\linewidth]{images/NGC1386-velMap.pdf}
	\end{subfigure}
	\begin{subfigure}{0.45\textwidth}
	\includegraphics[width=0.85\linewidth]{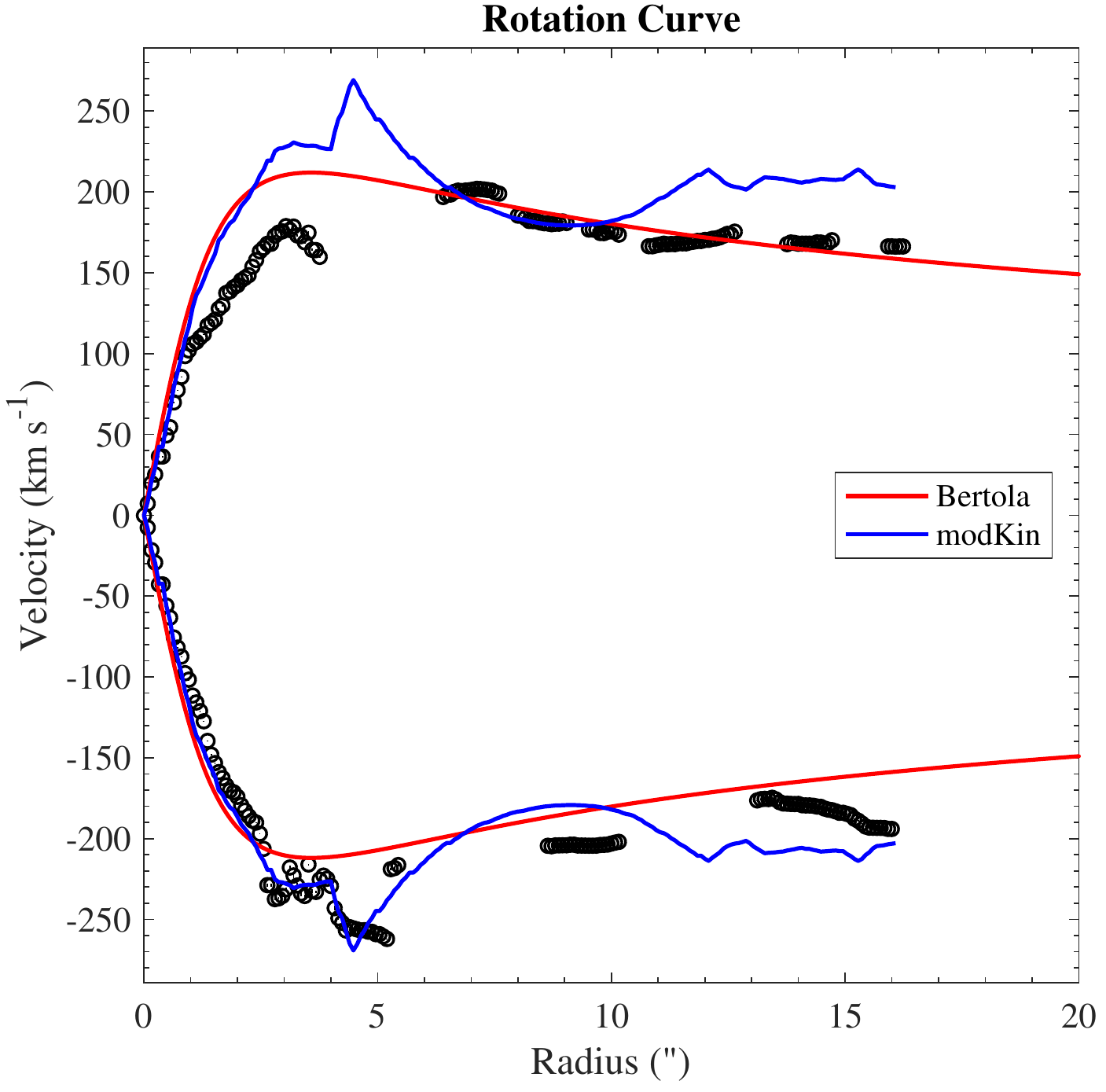}
	\end{subfigure}
	\begin{subfigure}{0.45\textwidth}
	\includegraphics[width=\linewidth]{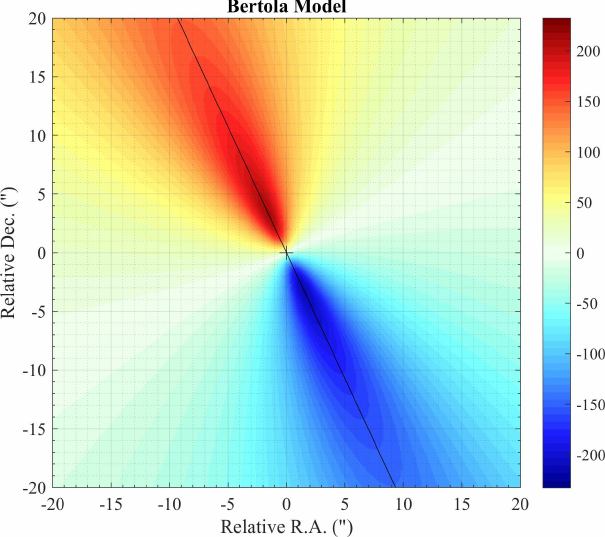}
	\end{subfigure}
	\begin{subfigure}{0.45\textwidth}
	\includegraphics[width=\linewidth]{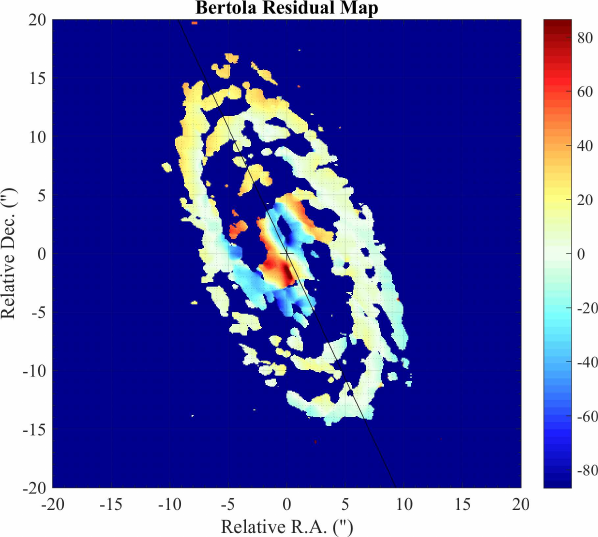}
	\end{subfigure}
	\begin{subfigure}{0.45\textwidth}
	\includegraphics[width=\linewidth]{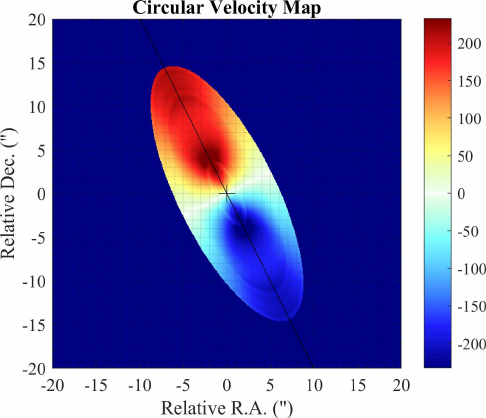}
	\end{subfigure}
	\begin{subfigure}{0.45\textwidth}
	\includegraphics[width=\linewidth]{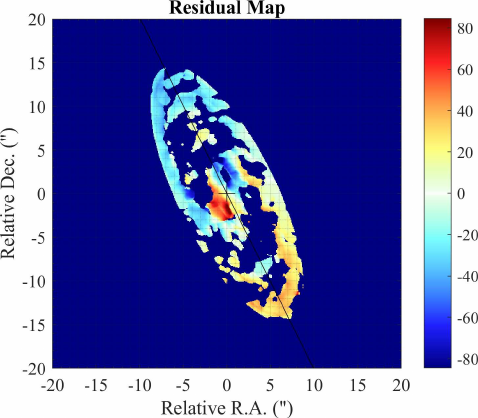}
	\end{subfigure}
	\caption{NGC~1386. \textit{Top-row:} ALMA velocity map (left) and rotation curves (right). \textit{Middle-row:} Bertola rotation model (left) and the corresponding residual map (right). \textit{Bottom-row:} Circular velocity model generated from the $B_{1}\cos\theta$ term of the \textit{modKin} package (left) and the residual map (right). The solid black line in all the images is the kinematic major axis. The black open circles in the rotation curve corresponds to the data while the red and the blue line corresponds to the rotation curve from Bertola and \textit{modKin} models that are shown in the left panel of the middle and bottom-row.}
	\label{NGC1386kin1}
\end{figure*}

In our attempt to model the kinematics of all the galaxies, we also attempted to constrain their kinematic centre. The models of all the galaxies and their residuals are shown in Figure~\ref{NGC1386kin1} and Appendix~\ref{kinMaps}. For both our models, we searched for the best-fitting kinematic centre within 5\arcsec\ of the continuum peak position. The result obtained from both the models seem to agree within a few ($\pm5$) pixels.  We decided to use the result from the \textit{modKin} code for further interpretation. The centre coordinates were compared to those given in 2MASS (Table~\ref{srcList}). For robustness, we consider the centre coordinates as a new estimate only if the difference in value is larger than the synthesized beam major axis. We, thus, were able to obtain new kinematic centres for NGC~1667 and NGC~5728. Interestingly, both these sources have a peculiar nuclear morphology as can be inferred from their CO moment 0 maps and a comparison of these with the HST images. NGC~1667 has two dominant CO-bright peaks in the nuclear region, separated by $\sim2$\arcsec\ (Figure~\ref{NGC1667pack}). The new kinematic center is coincident with the western CO-bright component while the 2MASS centre position is midway between the two CO-bright peaks. NGC~5728 is also characterised by two CO-bright components on the NE and SW part of the nucleus, separated by $\sim2$\arcsec. This galaxy is a candidate for a double rotating core \citep{Son2009}. In the HST maps (Figure~\ref{NGC5728pack}), the offset of the nucleus is quite evident, which could be a possible effect of a merger.
\begin{table}
	\centering
	\caption{New centre coordinates}
	\label{tabCent}
	\begin{tabular}{lccc}
	\toprule
	Sources     & R.A. & Dec & Separation \\
		    &      &     & (\arcsec)  \\
	\midrule
	NGC~1667    & 04:48:37.14 & -06:19:11.56 & 0.92 \\
	NGC~5728    & 14:42:23.87 & -17:15:10.93 & 0.94 \\
	\bottomrule
	\end{tabular}
\end{table}

\begin{figure*}
	\centering
	\begin{subfigure}[b]{0.48\textwidth}
		\includegraphics[width=\columnwidth]{images/NGC1386-velMap.pdf}
		\includegraphics[width=\columnwidth]{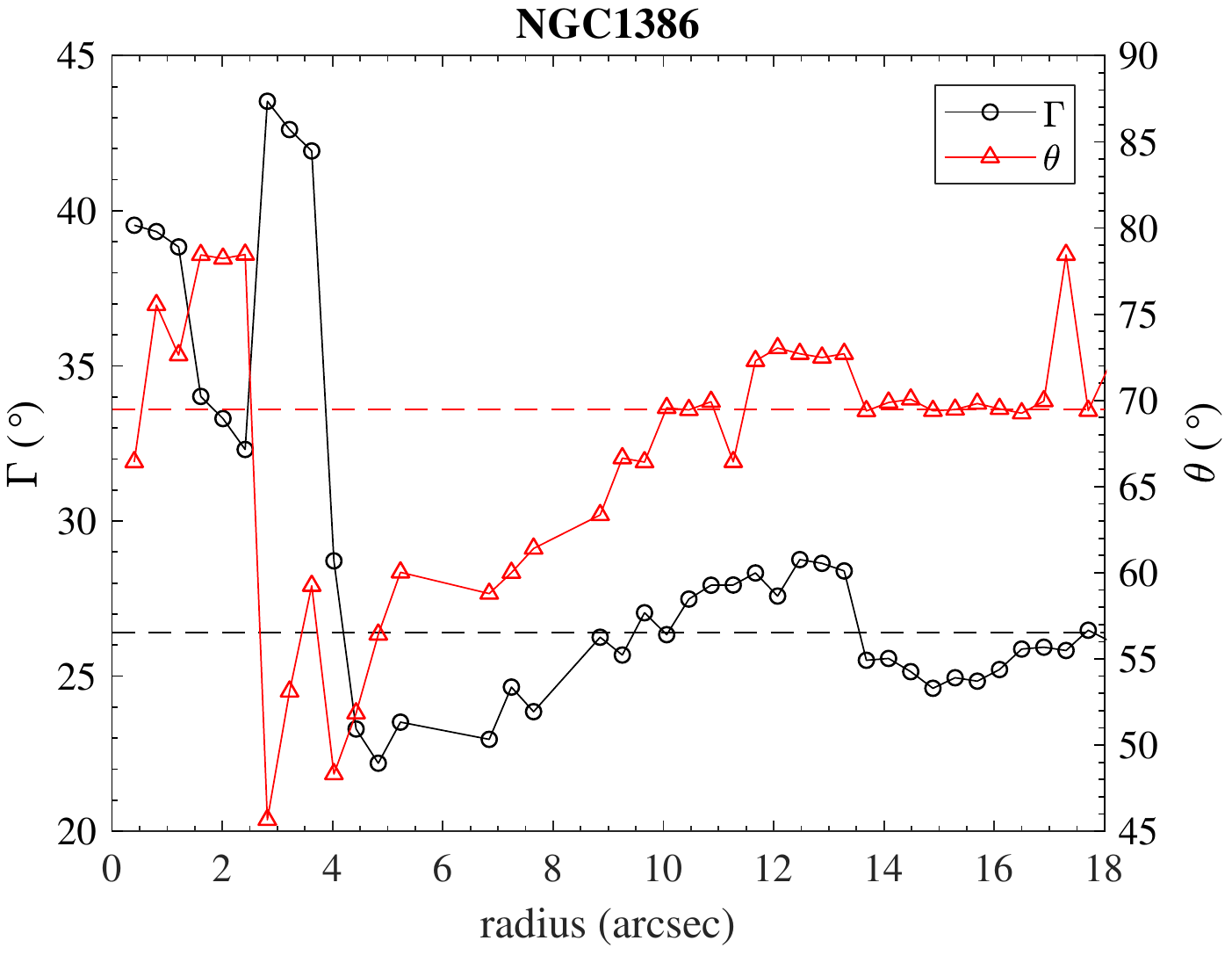}
	\end{subfigure}
	\begin{subfigure}[t]{0.48\textwidth}
		\includegraphics[width=\columnwidth]{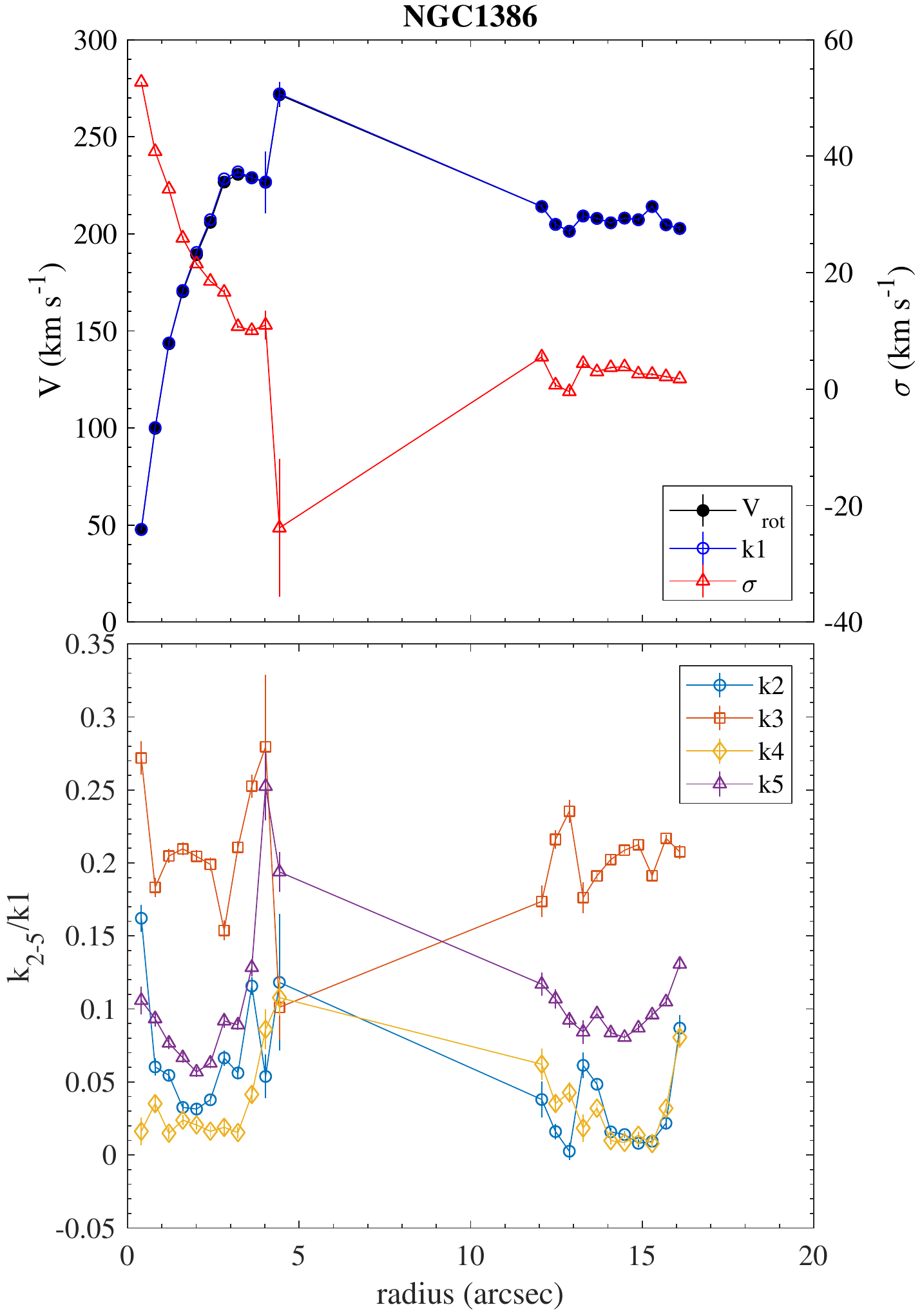}
	\end{subfigure}
	\caption{NGC1386. \textit{Top-row:} ALMA velocity map (left) and the velocity and velocity dispersion profile (right). The axis of the rotational velocity term $V_{\rm rot}$ and the Fourier amplitude term $k_1$ corresponds to the left ordinate while that for the velocity dispersion term ($\sigma$) corresponds to the right. \textit{Bottom-row:} Profile of the position angle and inclination along with their respective median values denoted by the dashed line (left) and the amplitudes of the Fourier coefficients (right). The left ordinate corresponds to the position angle and the right one to the inclination. All the amplitude of the Fourier coefficients that are plotted are normalised to the first term $k_1$.}
	\label{NGC1386kin2}
\end{figure*}

\subsection{Local kinematic results}

The radial dependence of the position angle ($\Gamma$), inclination ($\theta$) and the amplitude terms obtained from \textit{modKin} are vital in constraining the kinematic features in the velocity maps. The profiles for all the galaxies except for NGC~7213 (was impossible to obtain meaningful fits from \text{modKin} owing to the sparse information) are shown in the Figure~\ref{NGC1386kin2} and Appendix~\ref{kinMaps}. The radial profiles can be interpreted as follows:

\begin{itemize}
	\item The position angle in most galaxies show a smooth evolution with occasional variations, with the exception of NGC~5728 (Figure~\ref{NGC5728kin2}) which remains relatively constant for the most part, while a change of $\approx160^{\circ}$ is noticed in NGC~1667 (Figure~\ref{NGC1667kin2}).
	\item The change in inclination with radius of the galaxies ESO~428-G14, NGC~1386 and NGC~1667 seem relatively smooth; the occasional sudden fluctuations is most likely attributable to noise.
	\item The asymmetric circular velocity component (traced by the $B_1 \sin\theta$; hereafter denoted as $V_{\rm rot}$) and the $k_1$ amplitude term are both plotted in the figures so as to identify the galaxies with a departure from a pure circular rotation. Of the listed sources, NGC~1667 (Figure~\ref{NGC1667kin2}) and NGC~5728 (Figure~\ref{NGC5728kin2}) are the only ones that show a deviation from a circular rotation within the inner 2\arcsec. Otherwise, most galaxies show a smooth increase in $V_{\rm rot}$ which flattens at around 2\arcsec.
	\item There is growing evidence for power in even Fourier coefficients, despite the common practice of considering only the odd kinematic coefficients for the Moment 1 map, which could provide more clues to the perturbations in a galaxy. Hence, we show the amplitudes of terms $k_2, k_3, k_4$ and $k_5$ (normalised to the dominant term $k_1$ term). All galaxies show stark variations in all four amplitude parameters. This is discussed further in Section~\ref{instable}.
\end{itemize}

In works of SAURON \citep{Krajnovic2008} and ATLAS$^{\rm 3D}$ \citep{Krajnovic2011} surveys which studied the stellar kinematics in a large sample of galaxies, a classification scheme based on the profiles exhibited by parameters like the position angle, inclination and the amplitude terms of the Fourier coefficients, was introduced. All galaxies analysed here reveal multiple characteristics, such as a deviation greater than tens of degrees around the median value in position angle and inclination along with rapid variations in amplitude of most Fourier components. We adopt the same system as in the aforementioned works here to classify the galaxies as follows:

\begin{itemize}
	\item \textit{Kinematic twist:} Variation in position angle $>10^{\circ}$ around the median value. All sources with the exception of NGC~5728 and NGC~7213 fall into this category. These two galaxies show a smooth progression of the velocity as reflected by minor deviations of the position angle around the median value.
	\item \textit{Kinematically distinct core:} Abrupt variation of $>20^{\circ}$ in the position angle of adjacent rings along with the decline of the term $k_1$ to zero. In most cases, the ratio $k_5/k_1$ shows a peak around this region. NGC~1386 (Figure~\ref{NGC1386kin2}) is the only candidate that belongs to this class. The position angle of the source drops by $\sim20^{\circ}$ at 4\arcsec\ (220~pc) from the nucleus. Although $k_1$ does not show a significant decline, the $k_5/k_1$ term peaks around this radius.
	\item \textit{Counter-rotating core:} This is a unique case of the above category where the change in the position angle is of the order of $180^{\circ}$. NGC~1667 (Figure~\ref{NGC1667kin2}) exhibits such a property with a change of about $150^{\circ}$ in the inner 2\arcsec\ (634~pc) in addition to the significant drop in the rotational velocity term and a peak in $k_5/k_1$. The origin of such systems could be a result of a merger or an inflow of intergalactic clouds, in turn spurring the nuclear activity.
\end{itemize}

NGC~5728 (Figure~\ref{NGC5728kin2}) shows almost a constant phase in position angle. However, its inclination drops significantly ($\sim80^{\circ}$) in the inner 2--3\arcsec\ (428--642~pc) followed by a smooth increase. This trait and the decline of the rotational velocity in the same region favours the double rotating core hypothesis proposed by \citet{Son2009}. This feature will be explored in a subsequent work. We have restricted our discussion here to results obtained from considering only odd moments in the \textit{modKin} fits and note that any signatures such as those of possible mergers could have been constrained by modelling the even moments.

\subsection{Global results}

To understand the global kinematics of the galaxies we modelled the velocity maps using the Bertola model (Equation~\ref{BertolaEq}), which provides the maximum velocity, slope of the rotation curve after reaching its maximum and the parameter $c_0$, that gives the radius at which the velocity reaches 70\% of its maximum amplitude. This is obtained in addition to the position angle, inclination and the systemic velocity of the galaxy. The maximum velocity given by this model varies by $\sim450$~km~s$^{-1}$ across the galaxies. The slope of NGC~7213 is at the lower end (probably due to the sparsity in the data) while NGC~1667 occupies the upper bound. The slope of the remaining galaxies fall in the range $1.2-1.4$ (last column in Table~\ref{kinResults}). The region with the 70\% velocity in most systems are in the inner 1--2\arcsec\ with the exception of NGC~3081 (at 2.4\arcsec\ otherwise 463.2~pc) and ESO~428$-$G14 (at 4\arcsec\ otherwise 536~pc).

We compare the median values of the parameters -- position angle, inclination and systemic velocity -- obtained from the \textit{modKin} code with the Bertola model. These estimates are reported in columns~2 and 3 of Table~\ref{kinResults}. The position angle of all sources seems to comply within $\sim10^{\circ}$ of one another and the systemic velocity as well within $\sim20$~km~s$^{-1}$. Inclination seems to be the one with a relatively larger discrepancy, reaching as high as $\sim16^{\circ}$. In systems like ESO~428-G14 and NGC~1386, all three estimates seem to agree with each other. This inference does raise a claim for a disc-like rotation of these galaxies, which is indeed a favourable postulate from a visual inspection of their velocity maps.

The models under scrutiny in this work does generate a map of its own. Every such map is subtracted from the original velocity map of the galaxy to produce the residual, which provides information on the non-circular motion arising from various perturbations. The model used for this purpose in the case of \textit{modKin} is the one that is obtained from just the circular velocity term, therefore making the direct comparison of the residuals from both the models feasible. These models are shown in Figure~\ref{NGC1386kin1} and Appendix~\ref{kinMaps}. Amongst all the galaxies, the residual maps of NGC~5728 (Figure~\ref{NGC5728kin1}) are the only ones that are similar for the two models. In galaxies NGC~1667 (Figure~\ref{NGC1667kin1}), NGC~2110 (Figure~\ref{NGC2110kin1}) and NGC~3081 (Figure~\ref{NGC3081kin1}), the residuals reveal different possible perturbations. Finally, in ESO~428-G14 (Figure~\ref{ESO428kin1}) the residual maps in the outer region (beyond 3\arcsec otherwise $>$402~pc) and those in NGC~1386 (Figure~\ref{NGC1386kin2}) within the inner 3\arcsec\ ($<$165~pc) seem to be in agreement. The residual map of NGC~7213 (Figure~\ref{NGC7213kin1}) generated using the Bertola model reveals some perturbations in the inner 5\arcsec (550~pc). 

Apart from the maps mentioned above, we also extracted the rotation curves of all the galaxies along the major axis as given by the median of the position angle from the code \textit{modKin}. The rotation curves are plotted after the subtraction of the systemic velocity. Overplotted on every curve are those extracted from the two different models. NGC~5728 (Figure~\ref{NGC5728kin1}) is the only galaxy with a rotation curve suggesting a symmetric rotation as also supported by the models; NGC~1386 (Figure~\ref{NGC1386kin1}) seem to indicate a symmetric rotation for the most part with minor deviations in the inner 5\arcsec (275~pc). The observed rotation of the rest of the galaxies are either highly asymmetric or show strong deviations from the symmetry at various radii from the centre. No conclusion can be invoked from the rotation curve of NGC~7213.

There is clear evidence for the presence of non-axisymmetric features in certain galaxies, which for example, in NGC~1386 was also proven by CO (1-0) observations of \citet{Zabel2019}. We, however, caution the reader that the results reported in this work are primarily based on the interpretation of the intensity weighted mean velocity maps. As a result, we do not discuss non-axisymmetric features here; instead, these form the bulk of the discussion in a subsequent work (Ramakrishnan et~al. in prep.).

\subsection{Misalignment of photometric and kinematic axis}

Following \citet{Franx1991}, we compute the misalignment angle between the photometric and kinematic position angle as given by the following relation:
\begin{equation}
	\sin\Psi = \mid \sin({\rm PA_{phot}} - {\rm PA_{kin}}) \mid.
\end{equation}
The photometric position angle obtained from the 2MASS (Table~\ref{srcList}) is used in conjunction with the kinematic position angle reported in column~3 of Table~\ref{kinResults}. The misalignment angle thus obtained lies in the range $0-90^{\circ}$, regardless of the $180^{\circ}$ ambiguity between the two position angles. The estimated values are reported in Table~\ref{kinResults}. We classify all galaxies with $\Psi > 10^{\circ}$ as being misaligned. This makes ESO~428-G14, NGC~3081, NGC~5728 and NGC~7213 as misaligned candidates.

\section{Discussion and Summary}
\label{instable}

The correspondence between the molecular and ionised gas kinematics has been explored in earlier works \citep[e.g.][]{Wong2004, Davis2013a, deBlok2016} showing a good agreement in a majority of cases. In works by the ATLAS$^{\rm 3D}$ group, it was shown that the rotation velocities of the ionised gas to be much lower than that of the molecular ones. The circular rotation velocities obtained from the cold molecular gas are in general a better estimate over the stellar ones.

The close correspondence between the dust and the molecular gas of the seven galaxies can be inferred from the overlay of the moment 0 maps on the HST ones. This correspondence implies that the internal stellar mass loss is the primary factor for the production of the molecular gas.

All galaxies except NGC~5728 were already explored using the GEMINI IFU observations by our group \citep{Riffel2006, Lena2015, SchnorrMuller2014a, SchnorrMuller2014b, SchnorrMuller2016, SchnorrMuller2017a}. Throughout our study, we found the position angle of most galaxies, with the exception of NGC~1386, to be in agreement with those reported from the stellar kinematics (Table~\ref{kinResults}). The inclination and systemic velocity, however, seem to show a significant discrepancy. This disparity can be reconciled by an external origin scenario for the gas in such systems.

The local kinematics of the galaxies show trademarks of multicomponent systems. They were categorised into systems with kinematic twists, those with decoupled and a counter-rotating core. This classification already implies the prevalence of perturbations in the nuclear regions. The various perturbations that could play a vital role in these systems are outflow/inflow of gas, warping of spiral arms, barred (possibly multi-barred) systems and minor mergers, among others. Through the ionised gas kinematics, our group had already studied the outflows and inflows in various systems reported here. It is important to understand the effect of non-circular motions in these systems. The presence of such motions can be deciphered from the residual maps of various galaxies. The non-circular motions explain the dynamical processes in play and the transport of gas responsible under various phenomena under extreme conditions. The later scenario can be further understood by the fuelling processes that are governed either by external accretion or through secular process. We defer to the study of these motions and its connection to the ionised gas kinematics to a subsequent work (Ramakrishnan et al., in preparation).

We thus summarise our findings from this work as follows:

\begin{itemize}
		\item The CO integrated maps show spiral arms structure within the inner hundreds of parsecs with disc-like rotation.
		\item The continuum emission from three sources are point like while the other three show an extention varying from 3--6\arcsec. NGC~1667 is the only source with no continuum emission above the noise level.
		\item The local kinematics of the galaxies are explored using a modified version of the package \textit{Kinemetry}. Based on the radial profiles of the position angle, inclination and various amplitude terms, we are able to identify all galaxies as multi-component systems. They are further classified into those exhibiting a kinematic twist, with a decoupled core and also into a counter-rotating one. NGC~1667 is identified to fall into the latter category with the core counter-rotating by about $150^{\circ}$. The severity seen in the inclination profile of NGC~5728 seem to be an effect of two distinct cores as reported in \citet{Son2009}.
		\item The global kinematics are studied using the Bertola model, in addition to the median values of the local profile. The position angle and the systemic velocity agree within the errors, while the inclination shows a discrepancy. Using the global values, the misalignment angle is also computed showing that four out of seven galaxies to be misaligned with the photometric major axis.
		\item The centre position of two sources NGC~1667 and NGC~5728 are revised to a new estimate based on the kinematics. Both these sources are equally good candidates for a system with an offset nuclei.
		\item The residual maps of most galaxies show the presence of perturbations with clear signs of outflows/inflows in most systems. This will be a subject of future work.
\end{itemize}

\section*{ACKNOWLEDGEMENTS}

We thank the anonymous referee for valuable comments that improved this manuscript. VR acknowledges the support from the ALMA CONICYT project 31140007, 22930.921.02 Prog. Financ. BASAL-PFB/06 and Anillo ACT 172033. NN acknowledges support from BASAL-PFB/06 and extension, Fondecyt 1171506, and Anillo ACT 172033. RAR acknowledges support from CNPq and FAPERGS. This paper makes use of the following ALMA data: ADS/JAO.ALMA\#2012.1.00474.S and 2015.1.00086.S. ALMA is a partnership of ESO (representing its member states), NSF (USA) and NINS (Japan), together with NRC (Canada), MOST and ASIAA (Taiwan), and KASI (Republic of Korea), in cooperation with the Republic of Chile. The Joint ALMA Observatory is operated by ESO, AUI/NRAO and NAOJ. This research used the facilities of the Canadian Astronomy Data Centre operated by the National Research Council of Canada with the support of the Canadian Space Agency. This paper uses data taken with the NASA/ESA Hubble Space Telescope, obtained from the Data Archive at the Canadian Astronomy Data Centre. The Hubble Space Telescope is a collaboration between the Space Telescope Science Institute (STScI/NASA), the Space Telescope European Coordinating Facility (ST-ECF/ESA), and the Canadian Astronomy Data Centre (CADC/NRC/CSA). Use was also made of the NASA/IPAC Extragalactic Database (NED), which is operated by the Jet Propulsion Laboratory, California Institute of Technology, under contract with NASA. This research has made use of NASA's Astrophysics Data System. This research has made use of the SIMBAD database, operated at CDS, Strasbourg, France.

\bibliographystyle{mnras}
\bibliography{nStream}

\appendix

\section{Notes on Individual sources}
\label{sect2}

\textit{NGC~1386} is a Sb/c galaxy with a Seyfert 2 nucleus \citep{Malkan1998}. The HST imaging of this galaxy by \citet{Ferruit2000} reveals dusty nuclear region running parallel to the major axis. The ellipticity of the disc from the nucleus and outward increases from about 0.25 to 0.5 at $\sim1.5$~kpc with signs of warping. The radio observations of the source reveal a compact core with a jet that extends southward \citep{Nagar1999b}. Although a similar extension was reported in \oiii\ and \ha+\nii\ maps \citep{Weaver1991, StorchiBergmann1996, Ferruit2000}, no clear correspondence between the optical and radio was inferred \citep{Mundell2009}. The integral field spectroscopic observations in the optical by \citet{Lena2015} of the central $\sim500$~pc reveal a bright nuclear component with two lobes extending $\sim200$~pc north and south of the nucleus. Two kinematic features with velocity dispersions of $\approx90$ (entire field of view) and $\approx200$~km~s$^{-1}$ (inner 150~pc) was reported by the authors. 

\textit{NGC~1667} is a low-luminosity Sc galaxy with Seyfert 2 nucleus \citep{Malkan1998}. The galaxy is characterised by multiple spiral arms with dust contamination in the centrel kpc \citep{Schmitt2006}. Two roughly symmetrical spiral arms are visible in the nuclear region. The ultraviolet to IR observations of the galaxy by \citet{Schmitt2006} reveals that only a minor fraction of the UV emission originates from the nucleus unlike the \ha\ emission which is significantly stronger. Both the UV and \ha\ emission arise from the star-forming region along the spiral arms. The nuclear radio emission of the galaxy at 8.4~GHz is weak ($\sim7.5$~mJy) with no distinct features but a diffuse emission around the galaxy. The optical IFU observations by \citet{SchnorrMuller2017a} reveal complex morphology with two kinematic components of $\sigma \approx200$ and $\sigma \approx400$~km~s$^{-1}$ over the entire field of view and the inner 2\arcsec\ respectively.

\textit{NGC~2110} is a well-known Sa Seyfert 2 galaxy \citep{Malkan1998}. The HST \oiii and \ha+\nii\ emission line images of the galaxy is characterised by a dusty circumnuclear disc and a long S-shaped jet \citep{Mulchaey1994}. A similar jetlike feature was reported by \citet{Nagar1999b} that extends for about 4\arcsec\ along the north-south direction. Despite the prevalence of the extended emission along a similar position angle at both optical and radio wavebands, the emission regions are not causally-related. The peak of the radio emission is off-centred with no corresponding line emission. \citet{SchnorrMuller2014a} studied the feedback processes of the galaxy using the Gemini-IFU observations. The authors report four kinematic components of velocity dispersions in the range $60-220$~km~s$^{-1}$ for three of them, and a nuclear component in $220-600$~km~s$^{-1}$.

\textit{ESO~428-G14} is an S0/a Seyfert 2 galaxy as classified by \citet{deVaucouleurs1991}. The galaxy is characterised by a two-sided radio jet with an \oiii\ cone also being localised along the jet \citep{Falcke1996}. \oiii\ and \ha+\nii\ narrow-band images obtained with the HST show extended emission well aligned with the radio jet with stronger emission to the north-west as observed in the radio. The ratio \oiii/(\ha+\nii) shows a bipolar structure with larger values to the south-east \citep{Falcke1996, Falcke1998}. Non-circular motions based on the gas kinematics in the inner kiloparsec of the galaxy was reported by \citet{Riffel2006}. The galaxy major axis is noted to be aligned with the radio axis. From the radio emission morphology it is suggested that a radio jet is the cause of outflows.

\textit{NGC~3081} is an SB0/a galaxy hosting a Seyfert 2 nucleus \citep{Malkan1998}. The galaxy has a weak large-scale bar and a nuclear bar \citep{Buta1990} followed by four resonance rings -- two outer rings, an inner and a nuclear ring \citep{Buta1998}. Inside the nuclear ring, two spiral arms are observed. A comparison between the continuum adjacent to the \oiii\ and to the \ha+\nii\ lines showed that the nuclear ring and nuclear spiral arms are sites of recent or ongoing star formation \citep{Ferruit2000}. Radio observations revealed a compact radio source oriented roughly along the north-south direction \citep{Nagar1999b}. Two kinematic components were reported by \citet{SchnorrMuller2016} based on the optical IFU observations -- velocity dispersion of one in the range $\sigma\approx60-100$~km~s$^{-1}$ over the entire field of view and the other $\sigma\approx150-250$~km~s$^{-1}$ in the inner 200~pc.

\textit{NGC~5728} is a Seyfert 2 galaxy classified as SAB(r)a \citep{deVaucouleurs1991}. The HST \oiii\ and \ha+\nii\ narrow-band images by \citet{Wilson1993} show an extended biconical ionisation cone of $\sim1.8$~kpc in extent along the minor axis direction of the host galaxy. According to the kinematic analysis by \citet{Rubin1980}, the radial velocities of the central region could be interpreted as an expansion or rotation of the inner galactic disc region or a non-circular motion of a combination of both. There is an asymmetrical bright core at the centre surrounded by star-forming region in a ring of 5\arcsec\ in radius \citep{Schommer1988, Mazzuca2008}. The line spectra of the north-west region within 3\arcsec\ apart from the central nucleus showed double-peaked emission-line profiles \citep{Schommer1988}.

\textit{NGC~7213} is a Sa galaxy harbouring a Seyfert 1 AGN \citep{Malkan1998}. The galaxy has a bright nuclear point source with a very faint and featureless accretion disc. The galaxy is almost face on, hence giving a symmetrical appearance. The morphology of the galaxy is characterised by multiple spiral arms with severe dust extinction in the nuclear region. The 8.4~GHz VLA radio map of the galaxy shows an unresolved core \citep{Schmitt2001} with signs of variability \citep{Blank2005}. The stellar kinematics from the optical IFU studies by \citet{SchnorrMuller2014b} of this galaxy shows velocity dispersions of up to 200~km~s$^{-1}$ over the entire field of view.

\clearpage

\section{HST and ALMA images of galaxies}
\label{ALMAmaps}

\begin{figure*}
	\centering
	\begin{subfigure}{0.33\textwidth}
	\includegraphics[width=0.85\linewidth]{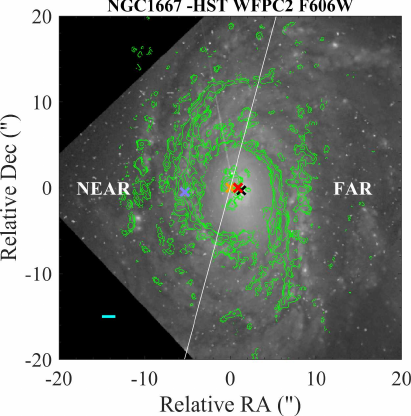}
	\end{subfigure}
	\begin{subfigure}{0.33\textwidth}
	\includegraphics[width=\linewidth]{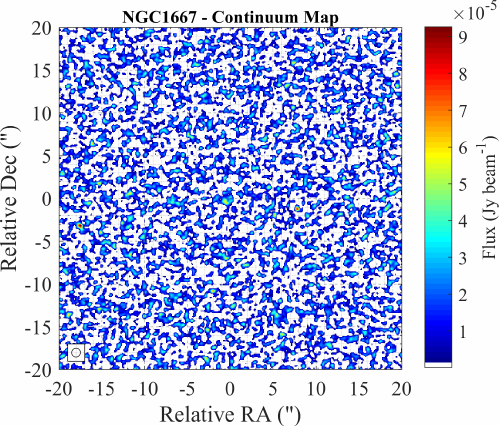}
	\end{subfigure}
	\begin{subfigure}{0.33\textwidth}
	\includegraphics[width=0.85\linewidth]{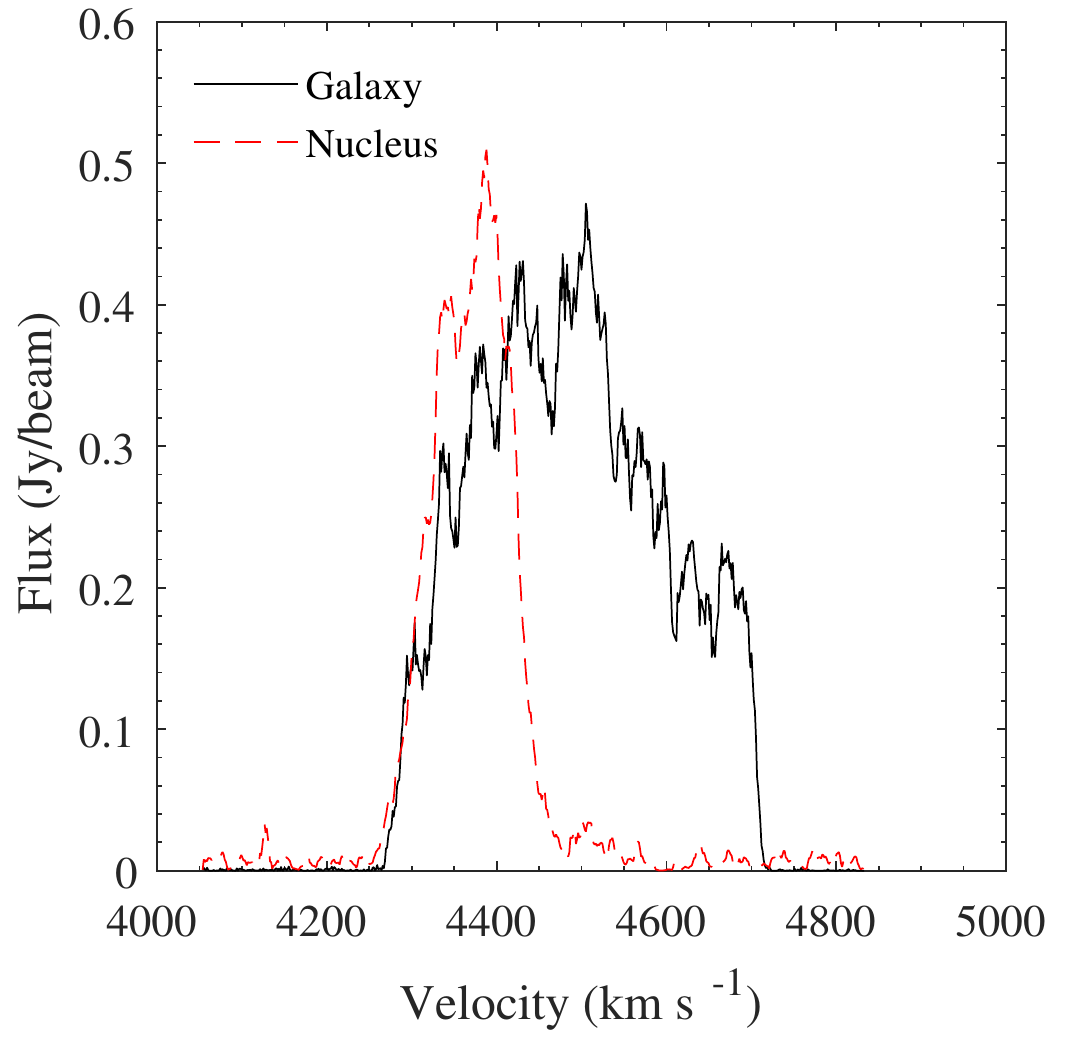}
	\end{subfigure}
	\begin{subfigure}{0.33\textwidth}
	\includegraphics[width=\linewidth]{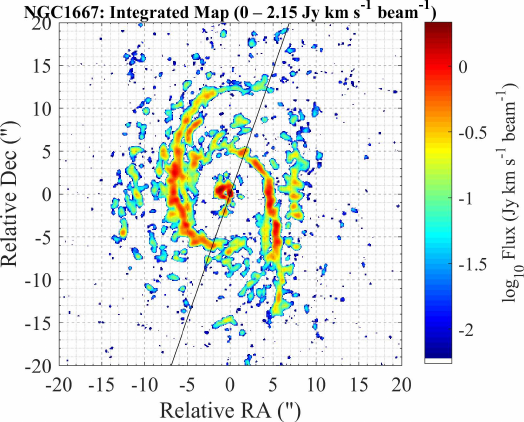}
	\end{subfigure}
	\begin{subfigure}{0.33\textwidth}
	\includegraphics[width=\linewidth]{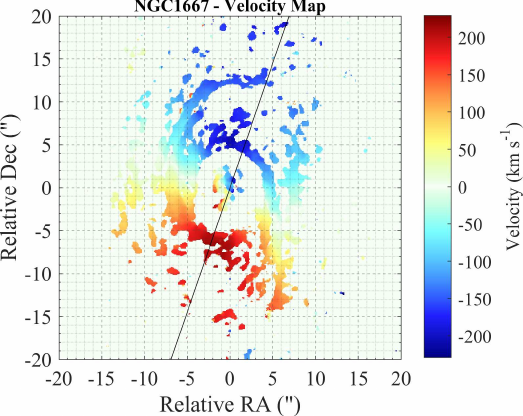}
	\end{subfigure}
	\begin{subfigure}{0.33\textwidth}
	\includegraphics[width=\linewidth]{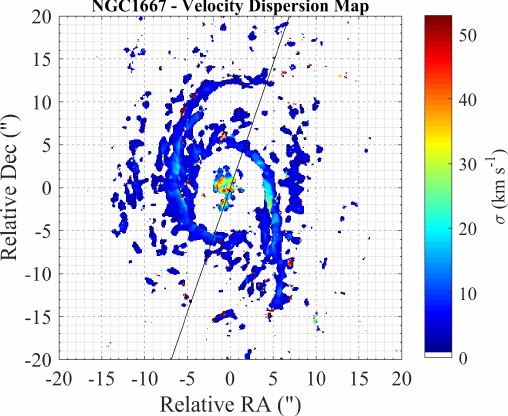}
	\end{subfigure}
	\caption{The same as in Figure~\ref{NGC1386pack} for NGC~1667.}
	\label{NGC1667pack}
\end{figure*}

\begin{figure*}
	\centering
	\begin{subfigure}{0.33\textwidth}
	\includegraphics[width=0.85\linewidth]{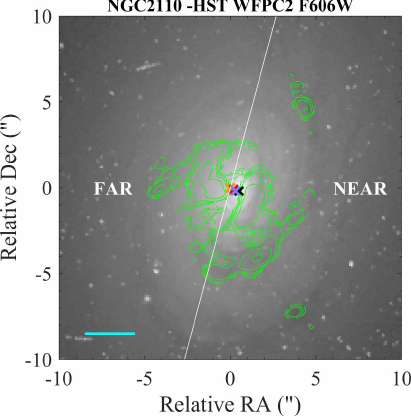}
	\end{subfigure}
	\begin{subfigure}{0.33\textwidth}
	\includegraphics[width=\linewidth]{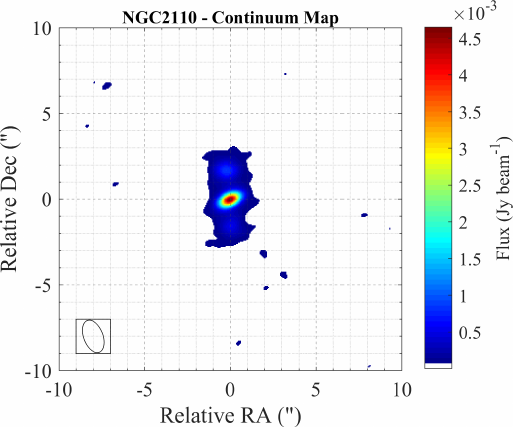}
	\end{subfigure}
	\begin{subfigure}{0.33\textwidth}
	\includegraphics[width=0.85\linewidth]{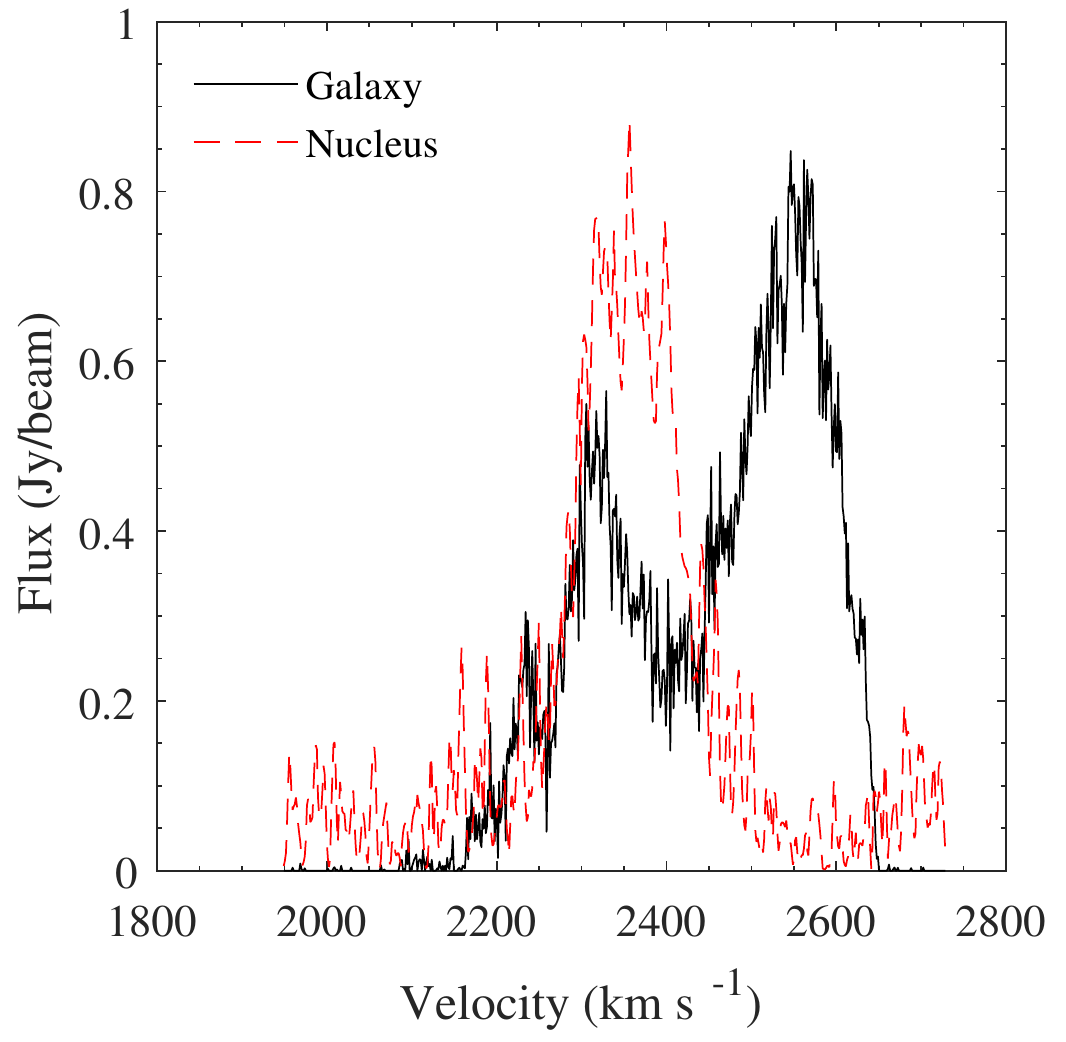}
	\end{subfigure}
	\begin{subfigure}{0.33\textwidth}
	\includegraphics[width=\linewidth]{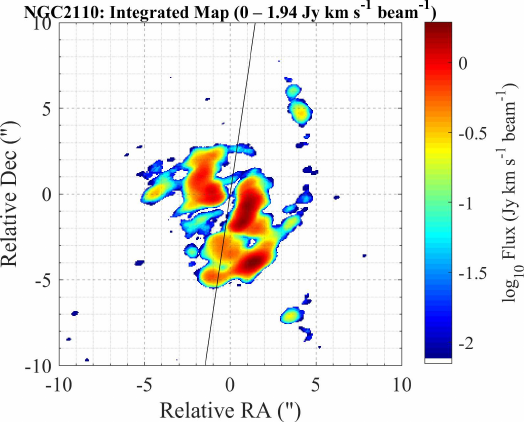}
	\end{subfigure}
	\begin{subfigure}{0.33\textwidth}
	\includegraphics[width=\linewidth]{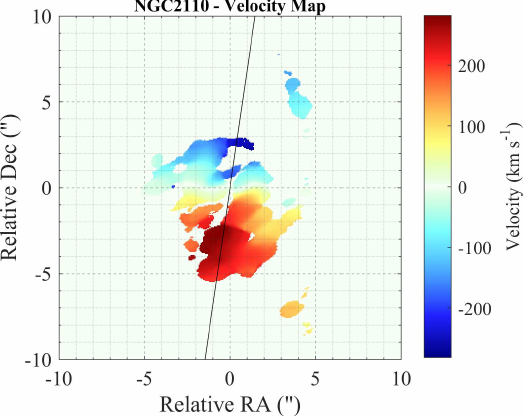}
	\end{subfigure}
	\begin{subfigure}{0.33\textwidth}
	\includegraphics[width=\linewidth]{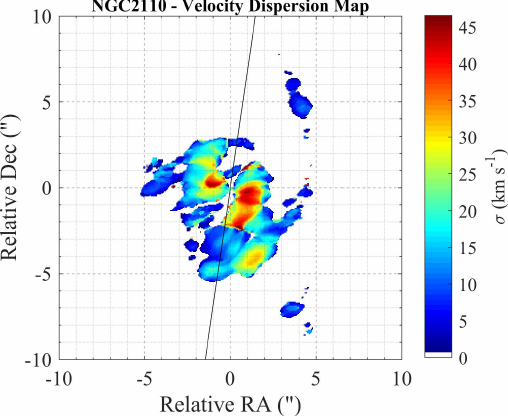}
	\end{subfigure}
	\caption{The same as in Figure~\ref{NGC1386pack} for NGC~2110.}
	\label{NGC2110pack}
\end{figure*}

\begin{figure*}
	\centering
	\begin{subfigure}{0.33\textwidth}
	\includegraphics[width=0.85\linewidth]{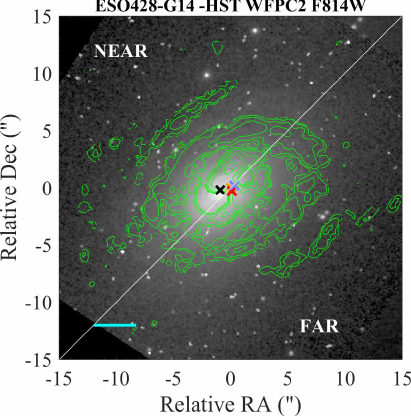}
	\end{subfigure}
	\begin{subfigure}{0.33\textwidth}
	\includegraphics[width=\linewidth]{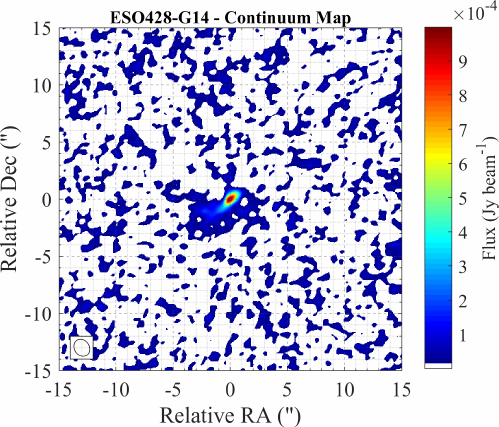}
	\end{subfigure}
	\begin{subfigure}{0.33\textwidth}
	\includegraphics[width=0.85\linewidth]{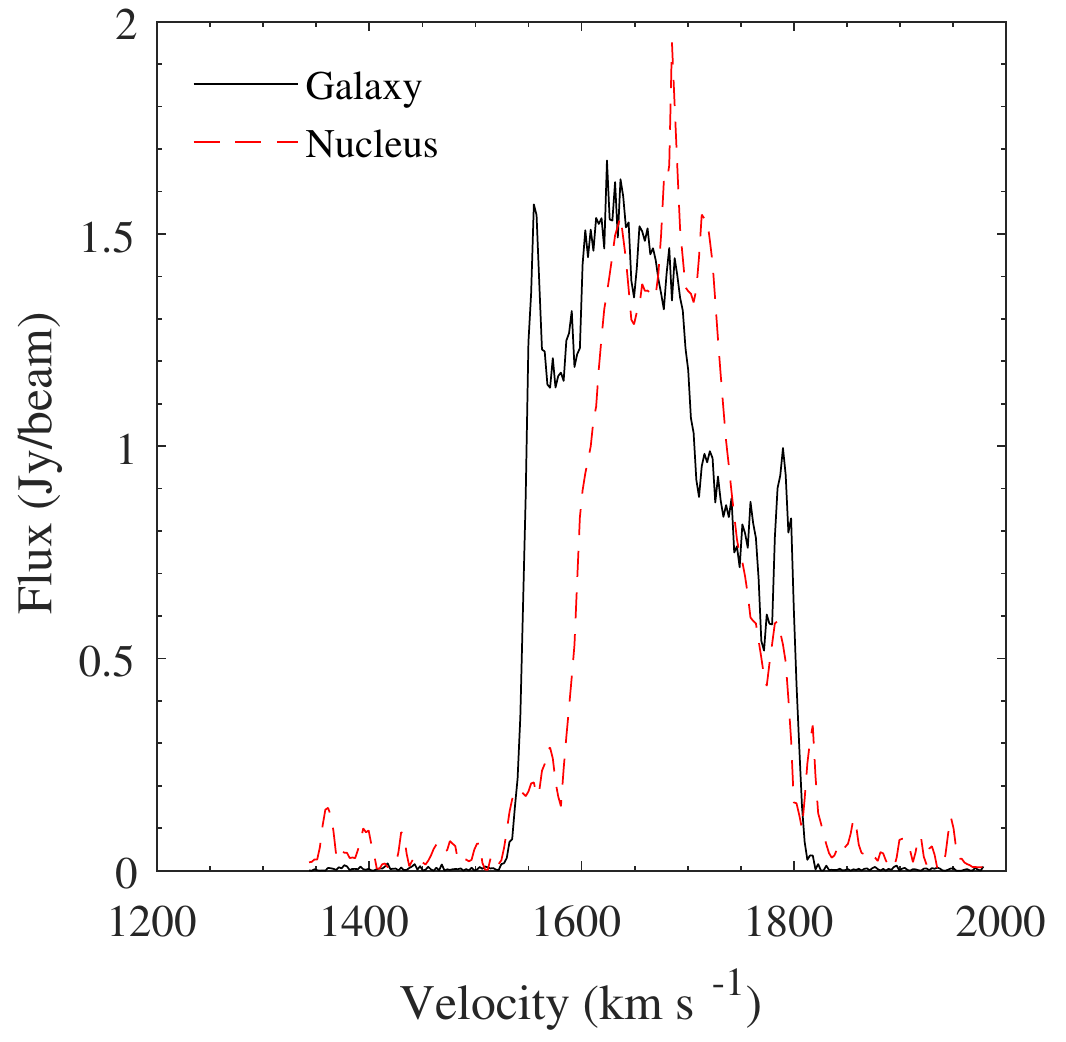}
	\end{subfigure}
	\begin{subfigure}{0.33\textwidth}
	\includegraphics[width=\linewidth]{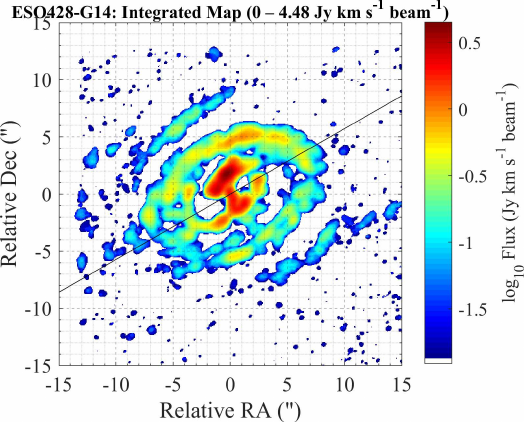}
	\end{subfigure}
	\begin{subfigure}{0.33\textwidth}
	\includegraphics[width=\linewidth]{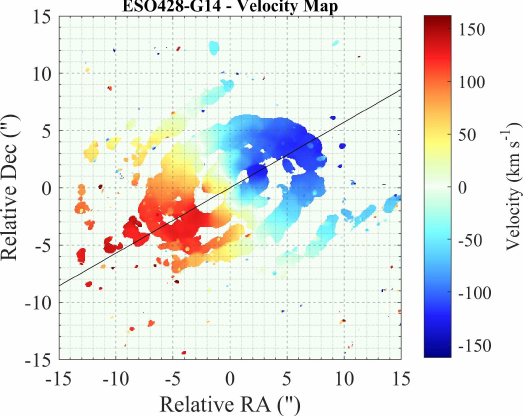}
	\end{subfigure}
	\begin{subfigure}{0.33\textwidth}
	\includegraphics[width=\linewidth]{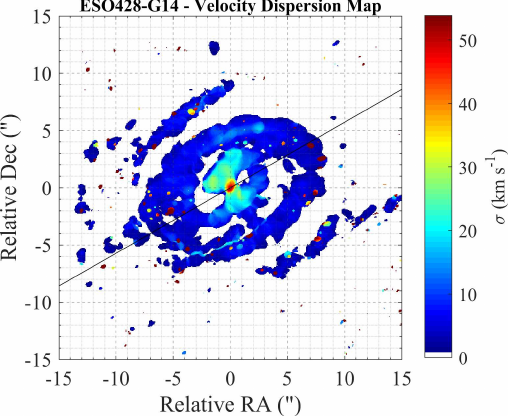}
	\end{subfigure}
	\caption{The same as in Figure~\ref{NGC1386pack} for ESO~428-G14.}
	\label{ESO428pack}
\end{figure*}

\begin{figure*}
	\centering
	\begin{subfigure}{0.33\textwidth}
	\includegraphics[width=0.85\linewidth]{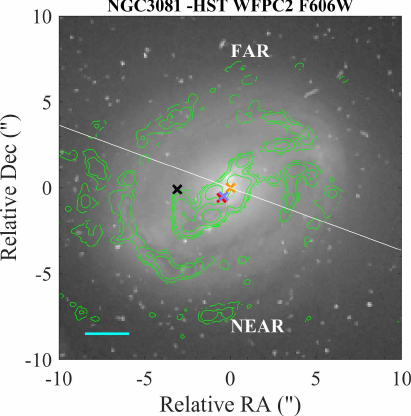}
	\end{subfigure}
	\begin{subfigure}{0.33\textwidth}
	\includegraphics[width=\linewidth]{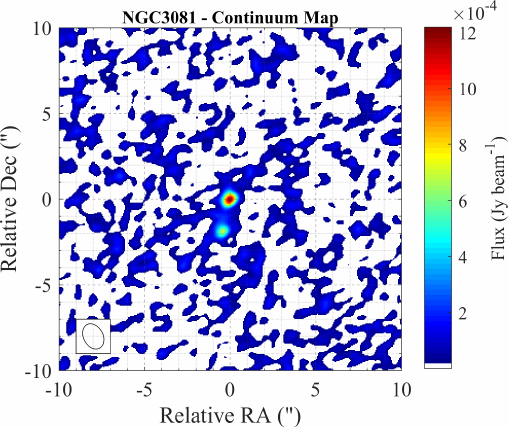}
	\end{subfigure}
	\begin{subfigure}{0.33\textwidth}
	\includegraphics[width=0.85\linewidth]{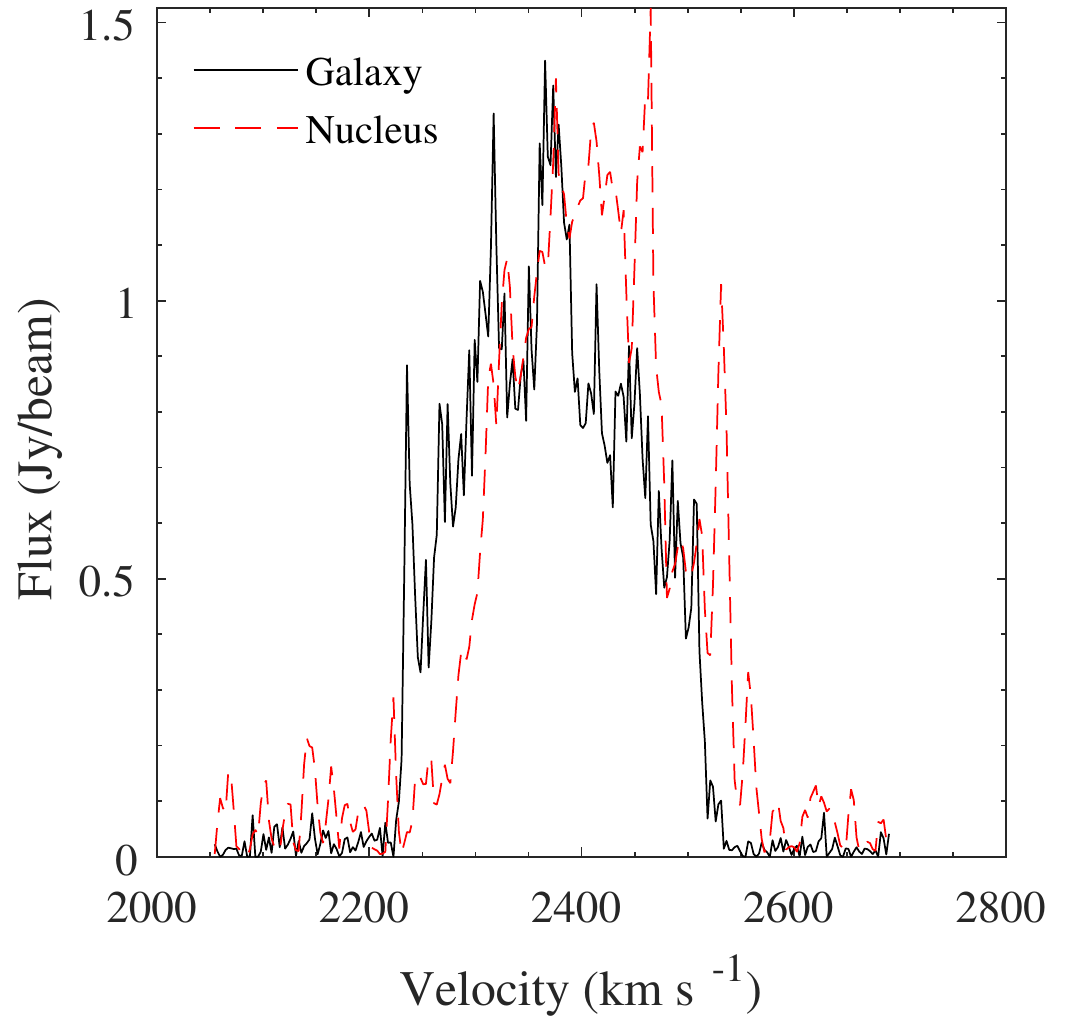}
	\end{subfigure}
	\begin{subfigure}{0.33\textwidth}
	\includegraphics[width=\linewidth]{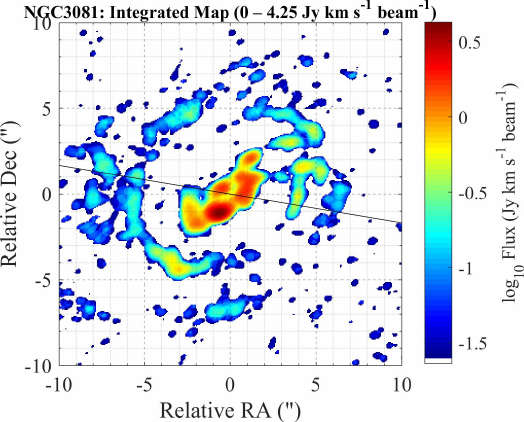}
	\end{subfigure}
	\begin{subfigure}{0.33\textwidth}
	\includegraphics[width=\linewidth]{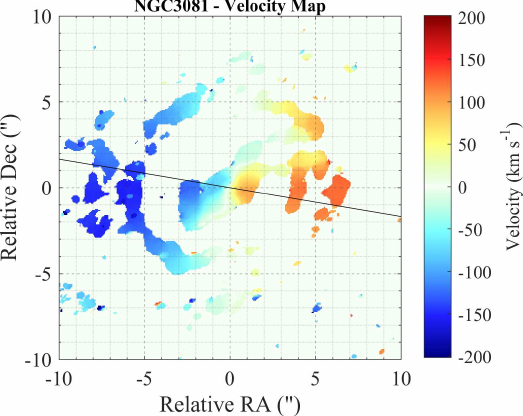}
	\end{subfigure}
	\begin{subfigure}{0.33\textwidth}
	\includegraphics[width=\linewidth]{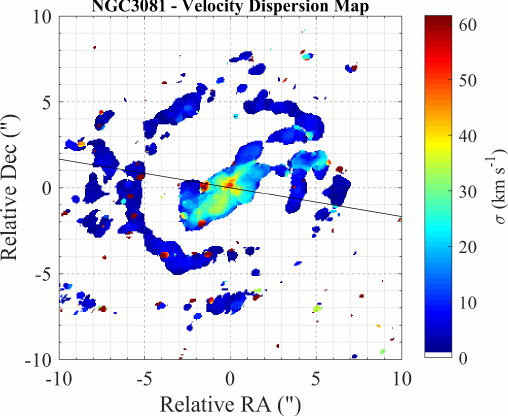}
	\end{subfigure}
	\caption{The same as in Figure~\ref{NGC1386pack} for NGC~3081.}
	\label{NGC3081pack}
\end{figure*}

\begin{figure*}
	\centering
	\begin{subfigure}{0.33\textwidth}
	\includegraphics[width=0.85\linewidth]{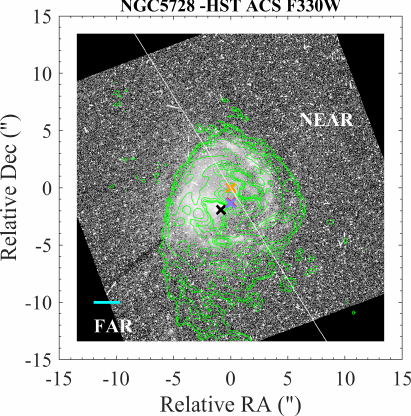}
	\end{subfigure}
	\begin{subfigure}{0.33\textwidth}
	\includegraphics[width=\linewidth]{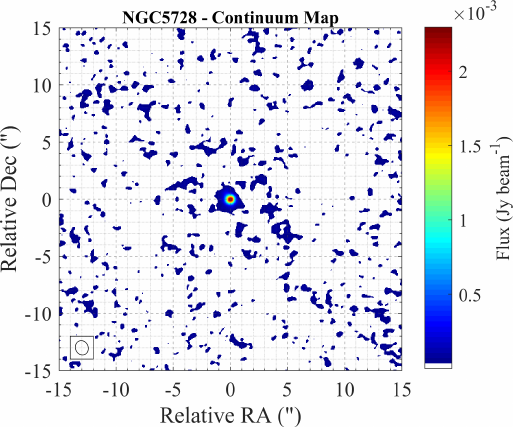}
	\end{subfigure}
	\begin{subfigure}{0.33\textwidth}
	\includegraphics[width=0.85\linewidth]{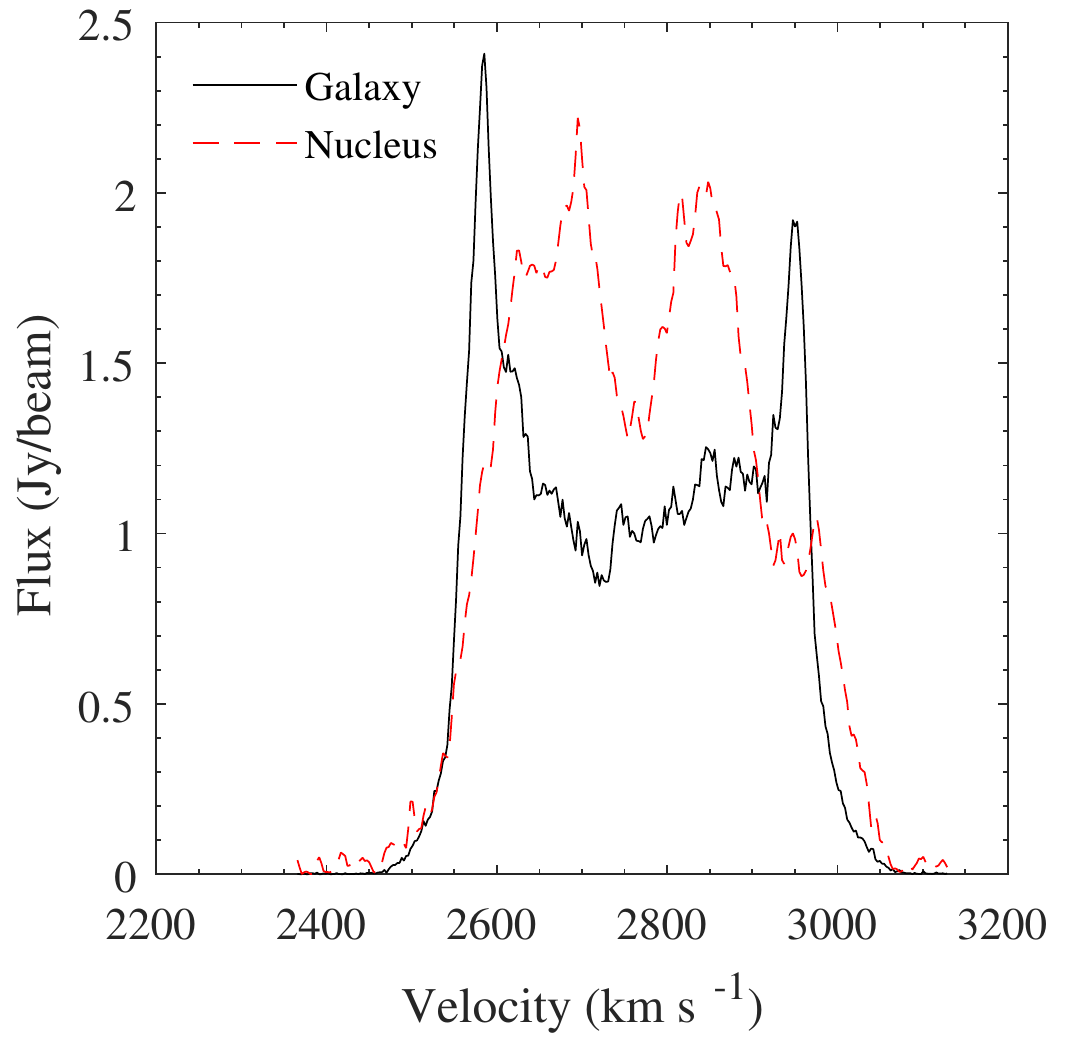}
	\end{subfigure}
	\begin{subfigure}{0.33\textwidth}
	\includegraphics[width=\linewidth]{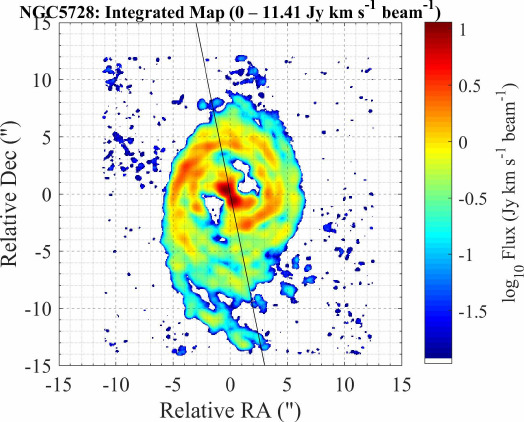}
	\end{subfigure}
	\begin{subfigure}{0.33\textwidth}
	\includegraphics[width=\linewidth]{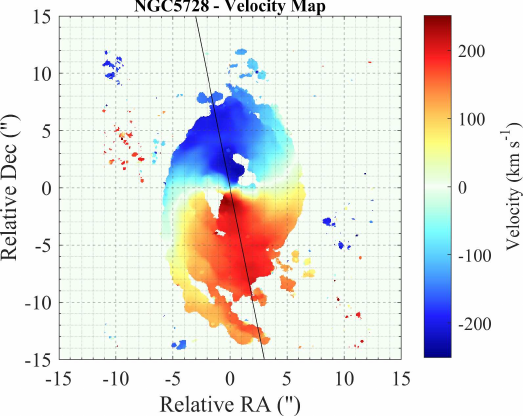}
	\end{subfigure}
	\begin{subfigure}{0.33\textwidth}
	\includegraphics[width=\linewidth]{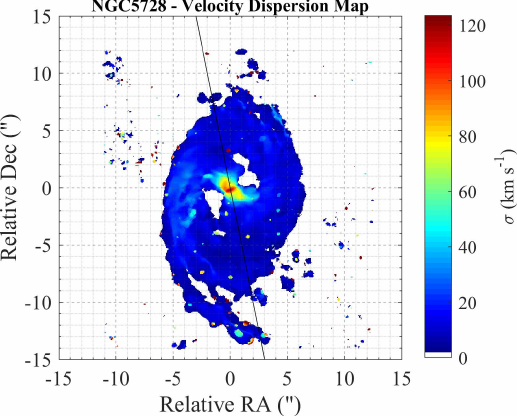}
	\end{subfigure}
	\caption{The same as in Figure~\ref{NGC1386pack} for NGC~5728.}
	\label{NGC5728pack}
\end{figure*}

\begin{figure*}
	\centering
	\begin{subfigure}{0.33\textwidth}
	\includegraphics[width=0.85\linewidth]{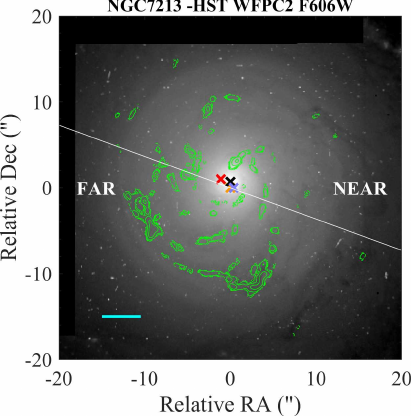}
	\end{subfigure}
	\begin{subfigure}{0.33\textwidth}
	\includegraphics[width=\linewidth]{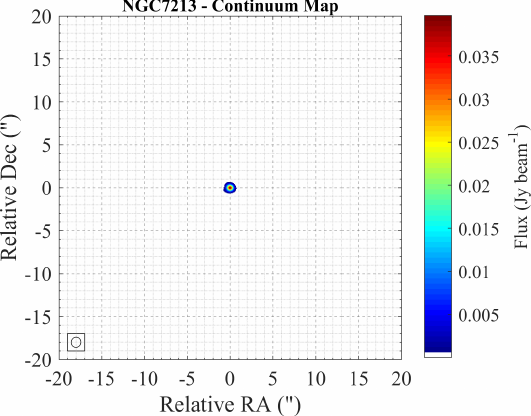}
	\end{subfigure}
	\begin{subfigure}{0.33\textwidth}
	\includegraphics[width=0.85\linewidth]{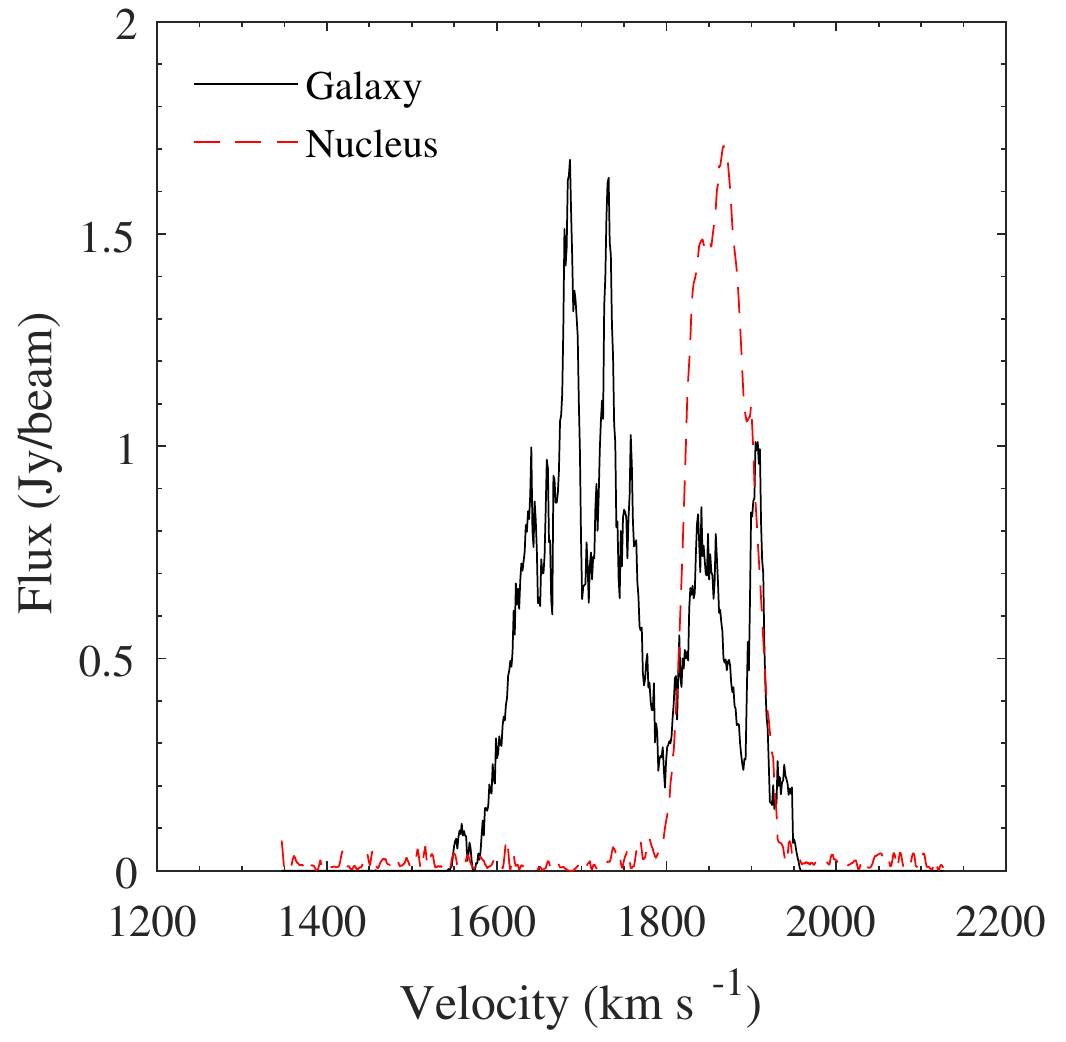}
	\end{subfigure}
	\begin{subfigure}{0.33\textwidth}
	\includegraphics[width=\linewidth]{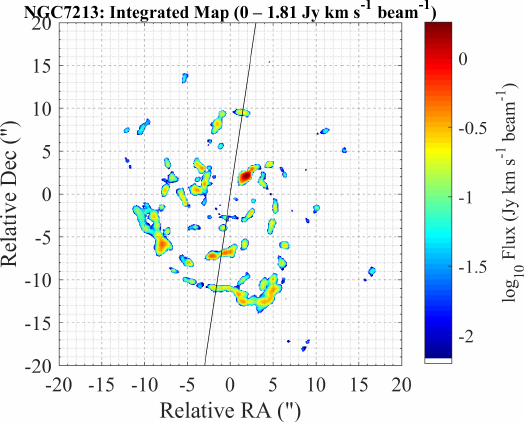}
	\end{subfigure}
	\begin{subfigure}{0.33\textwidth}
	\includegraphics[width=\linewidth]{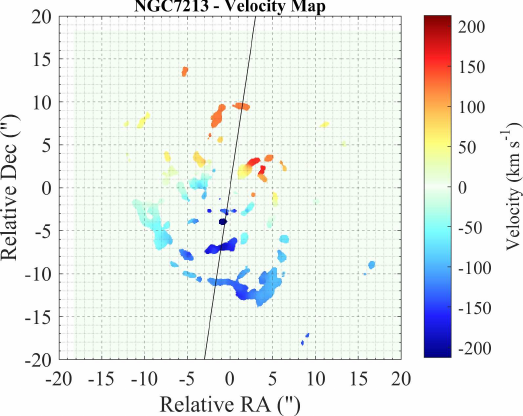}
	\end{subfigure}
	\begin{subfigure}{0.33\textwidth}
	\includegraphics[width=\linewidth]{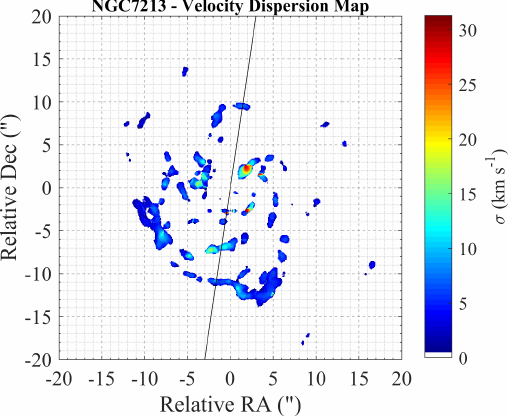}
	\end{subfigure}
	\caption{The same as in Figure~\ref{NGC1386pack} for NGC~7213.}
	\label{NGC7213pack}
\end{figure*}

\clearpage

\section{Results from Kinematic studies}
\label{kinMaps}

\begin{figure*}
	\centering
	\begin{subfigure}{0.45\textwidth}
	\includegraphics[width=\linewidth]{images/NGC1667-velMap.pdf}
	\end{subfigure}
	\begin{subfigure}{0.45\textwidth}
	\includegraphics[width=0.85\linewidth]{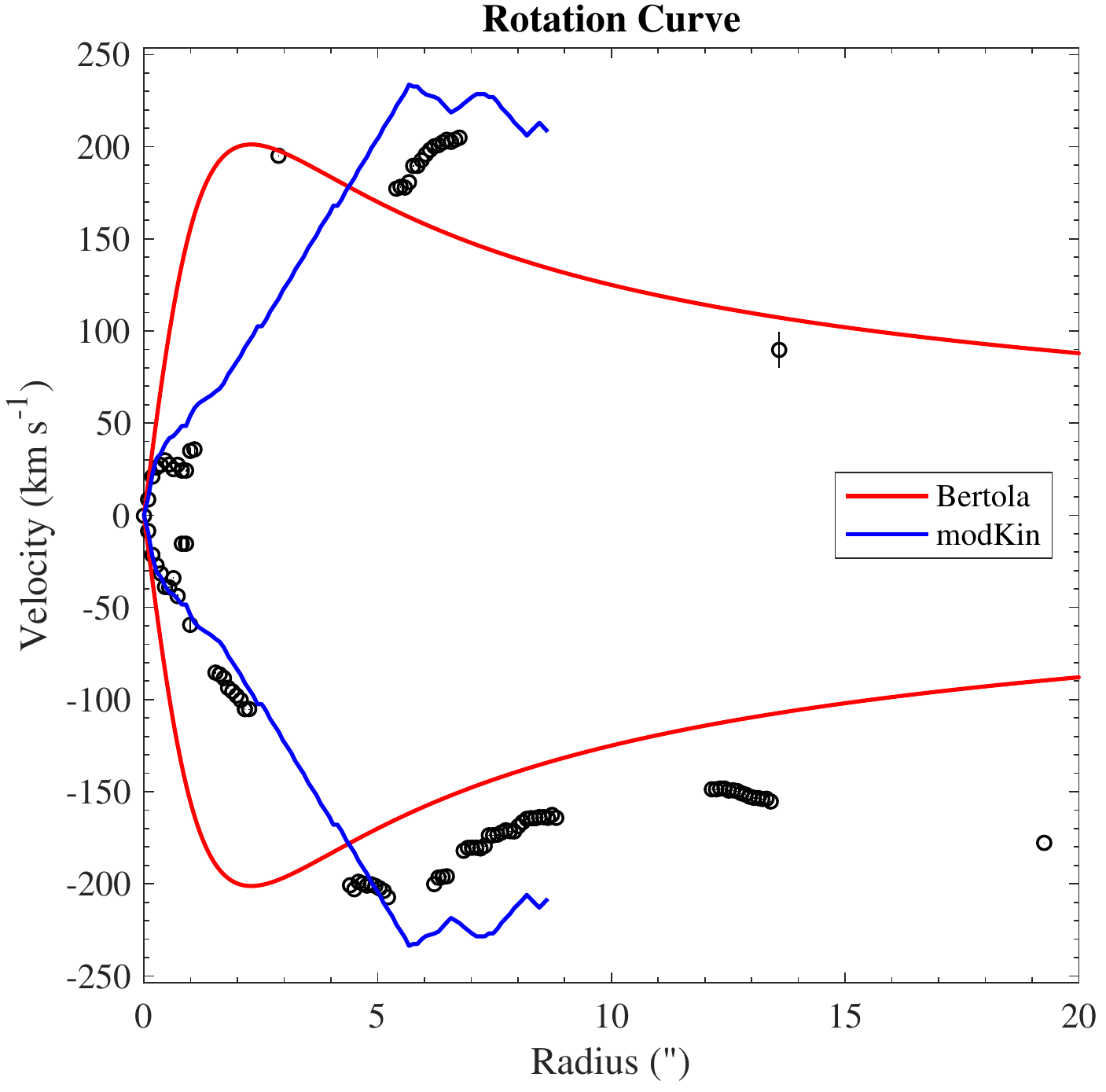}
	\end{subfigure}
	\begin{subfigure}{0.45\textwidth}
	\includegraphics[width=\linewidth]{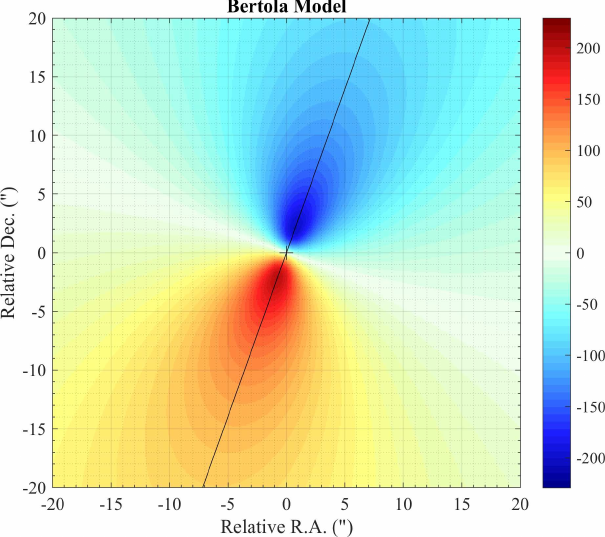}
	\end{subfigure}
	\begin{subfigure}{0.45\textwidth}
	\includegraphics[width=\linewidth]{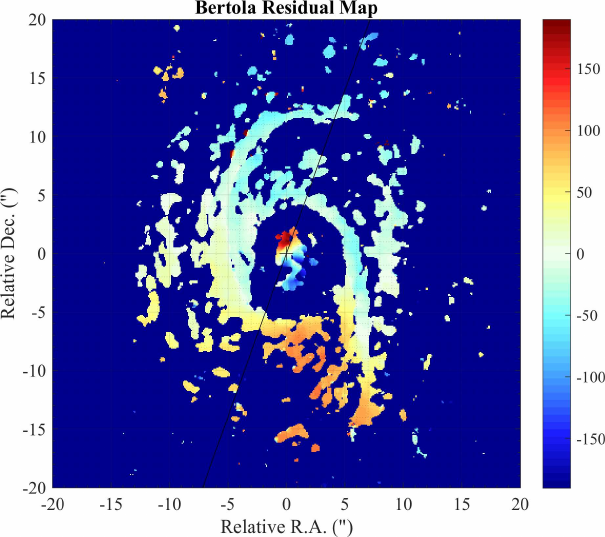}
	\end{subfigure}
	\begin{subfigure}{0.45\textwidth}
	\includegraphics[width=\linewidth]{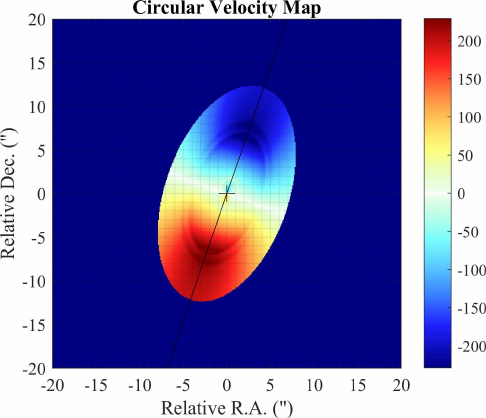}
	\end{subfigure}
	\begin{subfigure}{0.45\textwidth}
	\includegraphics[width=\linewidth]{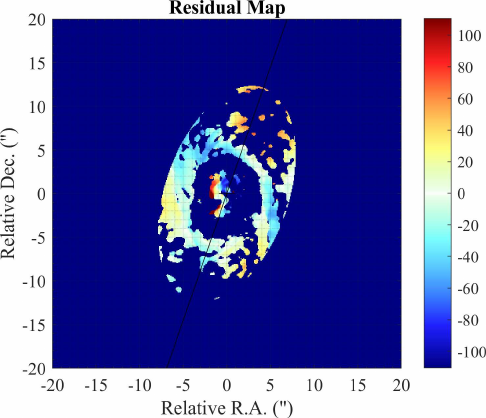}
	\end{subfigure}
	\caption{The same as in Figure~\ref{NGC1386kin1} for NGC~1667.}
	\label{NGC1667kin1}
\end{figure*}

\begin{figure*}
	\centering
	\begin{subfigure}[b]{0.48\textwidth}
		\includegraphics[width=\columnwidth]{images/NGC1667-velMap.pdf}
		\includegraphics[width=\columnwidth]{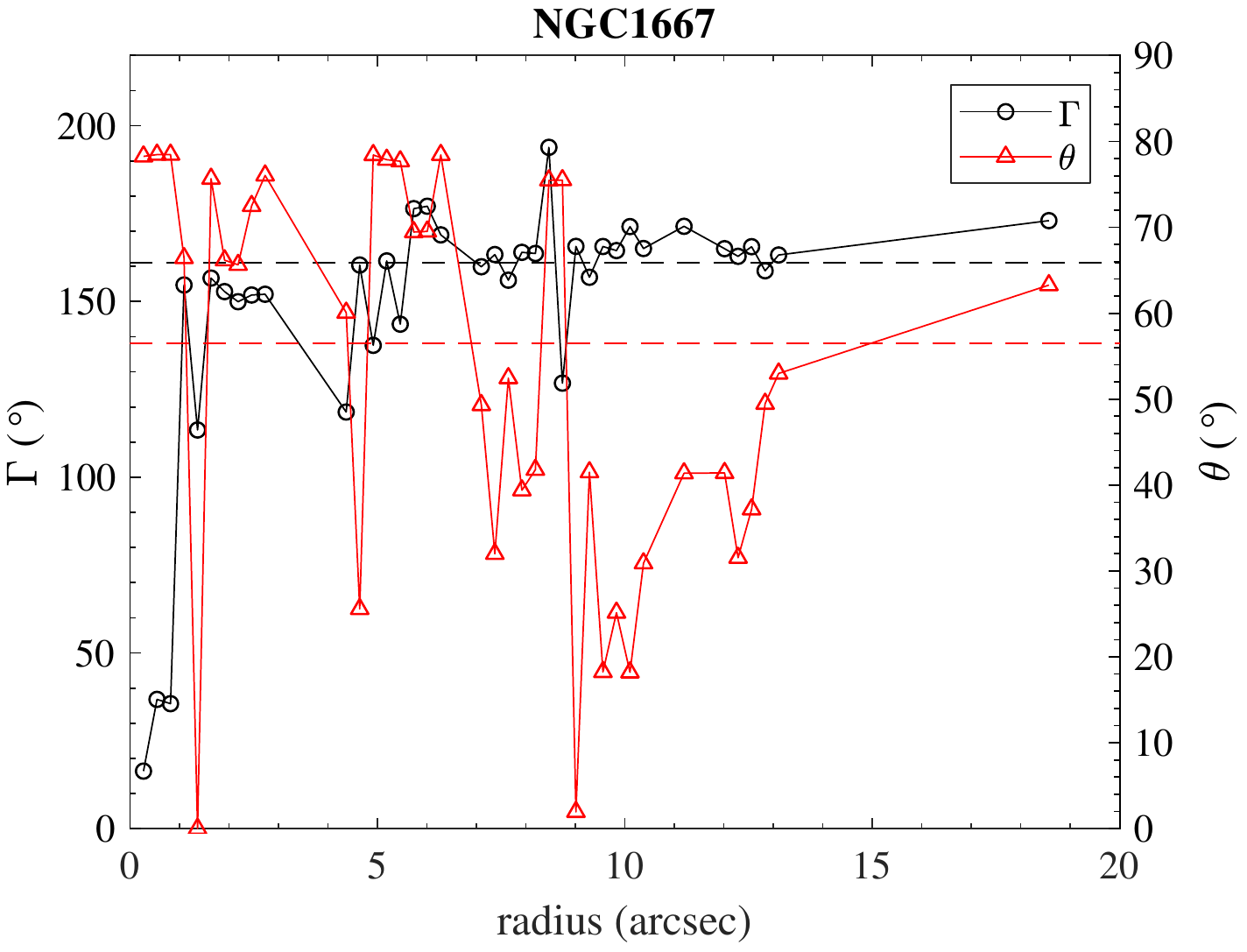}
	\end{subfigure}
	\begin{subfigure}[t]{0.48\textwidth}
		\includegraphics[width=\columnwidth]{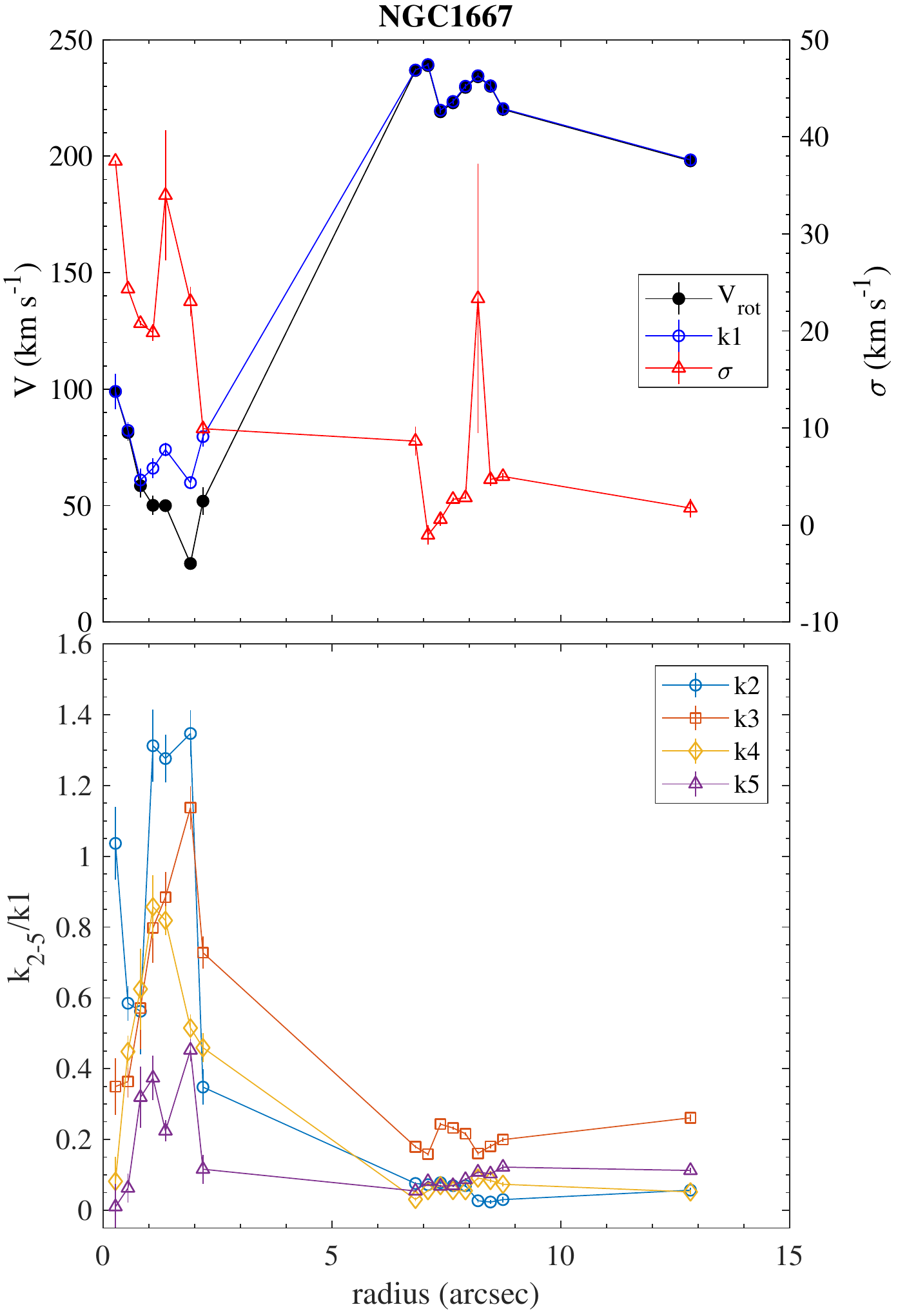}
	\end{subfigure}
	\caption{The same as in Figure~\ref{NGC1386kin2} for NGC~1667.}
	\label{NGC1667kin2}
\end{figure*}

\begin{figure*}
	\centering
	\begin{subfigure}{0.45\textwidth}
	\includegraphics[width=\linewidth]{images/NGC2110-velMap.pdf}
	\end{subfigure}
	\begin{subfigure}{0.45\textwidth}
	\includegraphics[width=0.85\linewidth]{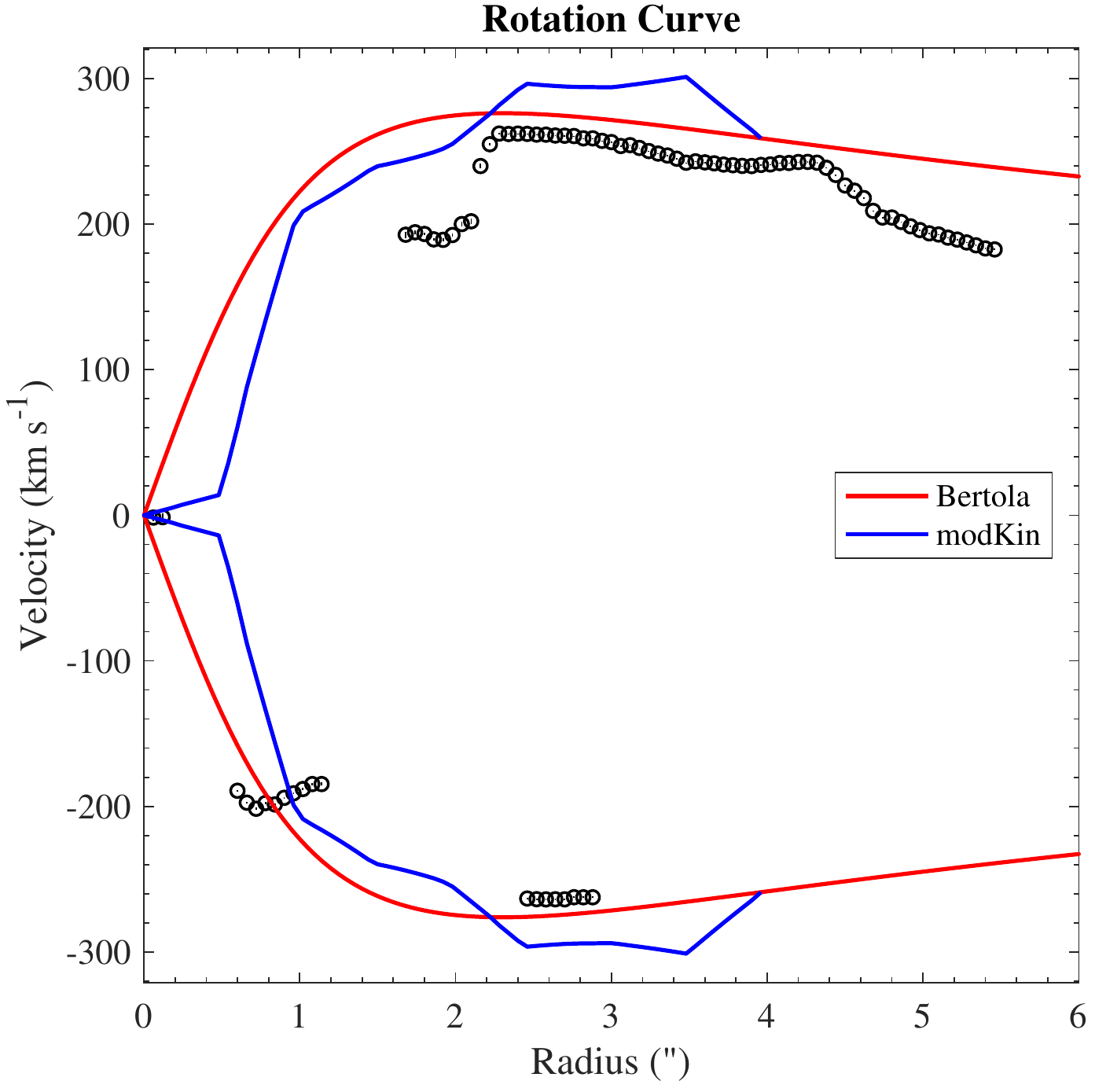}
	\end{subfigure}
	\begin{subfigure}{0.45\textwidth}
	\includegraphics[width=\linewidth]{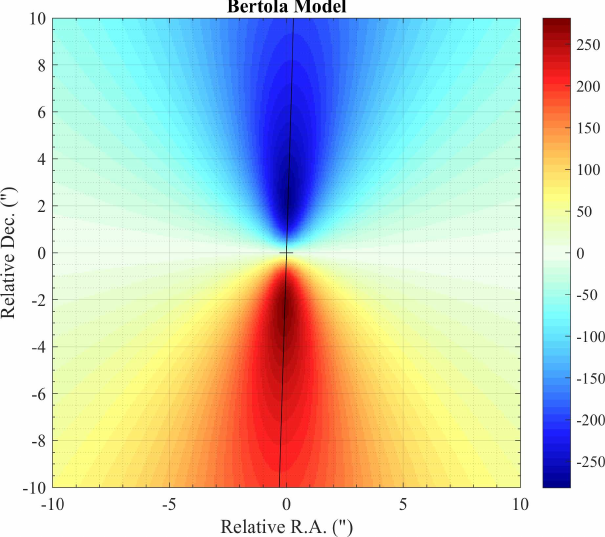}
	\end{subfigure}
	\begin{subfigure}{0.45\textwidth}
	\includegraphics[width=\linewidth]{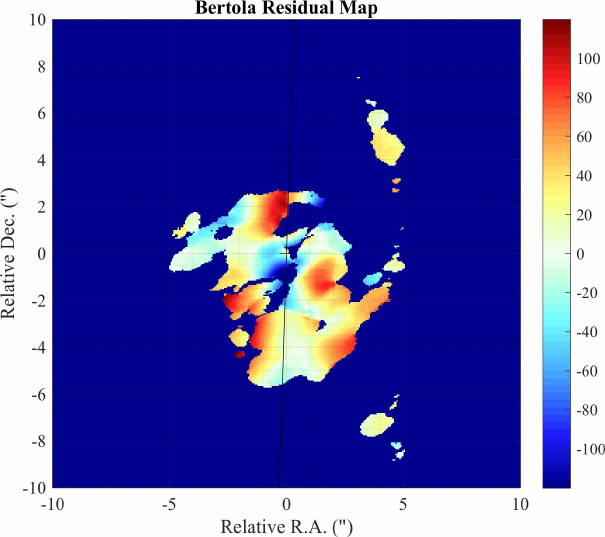}
	\end{subfigure}
	\begin{subfigure}{0.45\textwidth}
	\includegraphics[width=\linewidth]{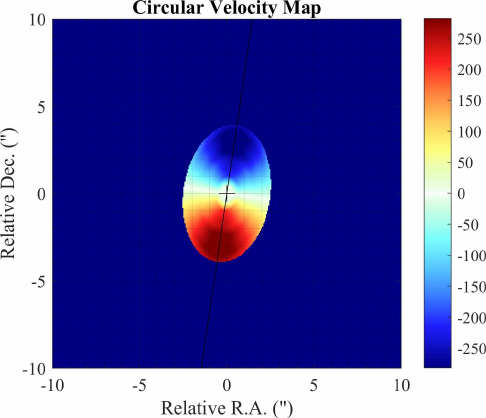}
	\end{subfigure}
	\begin{subfigure}{0.45\textwidth}
	\includegraphics[width=\linewidth]{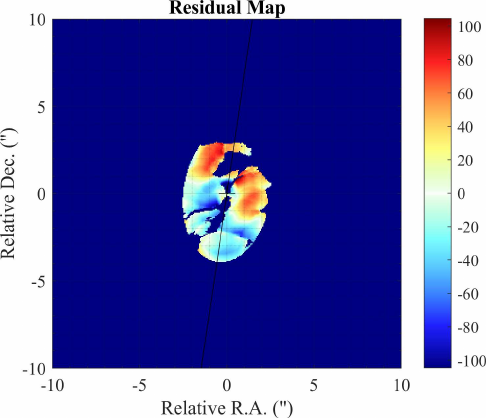}
	\end{subfigure}
	\caption{The same as in Figure~\ref{NGC1386kin1} for NGC~2110.}
	\label{NGC2110kin1}
\end{figure*}

\begin{figure*}
	\centering
	\begin{subfigure}[b]{0.48\textwidth}
		\includegraphics[width=\columnwidth]{images/NGC2110-velMap.pdf}
		\includegraphics[width=\columnwidth]{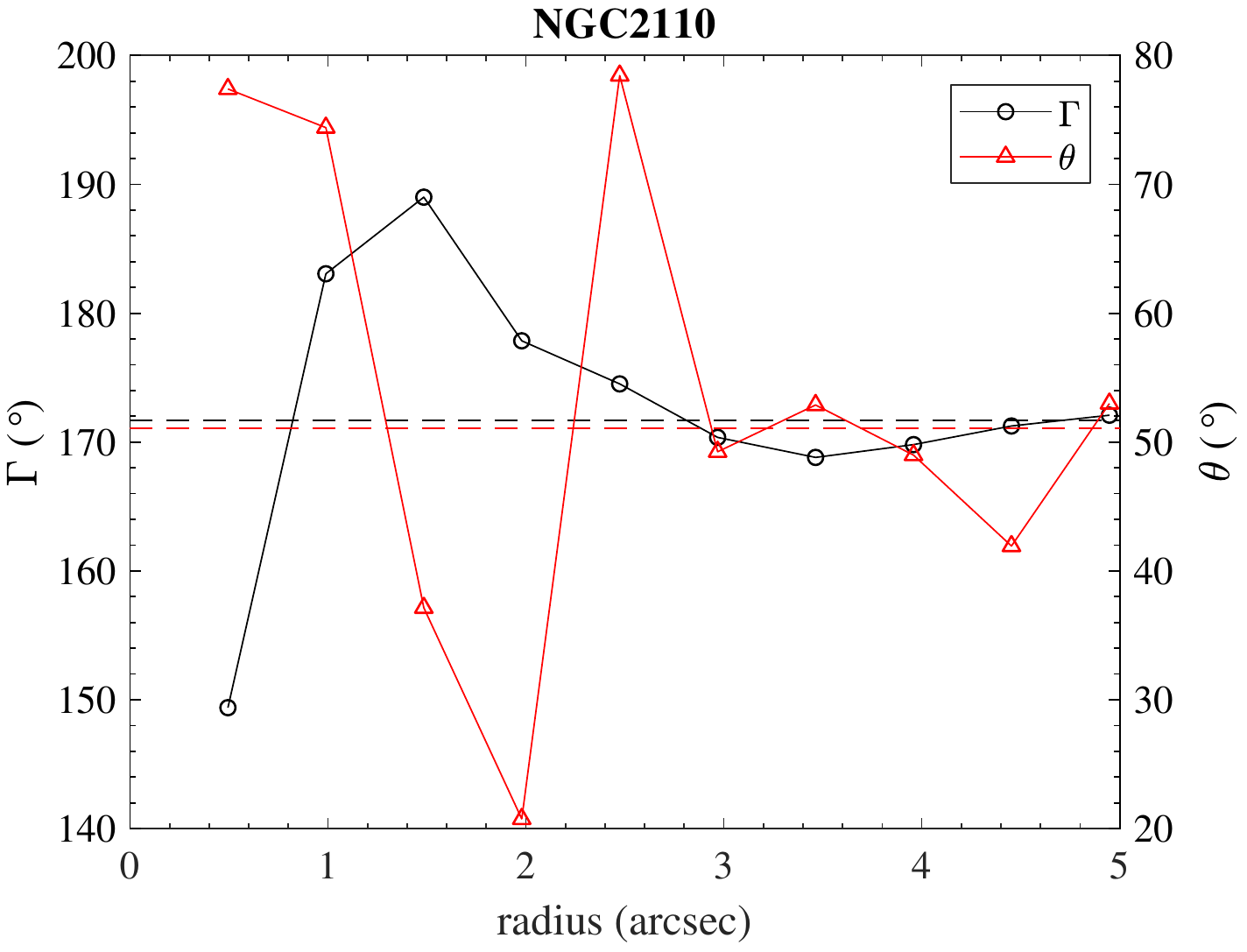}
	\end{subfigure}
	\begin{subfigure}[t]{0.48\textwidth}
		\includegraphics[width=\columnwidth]{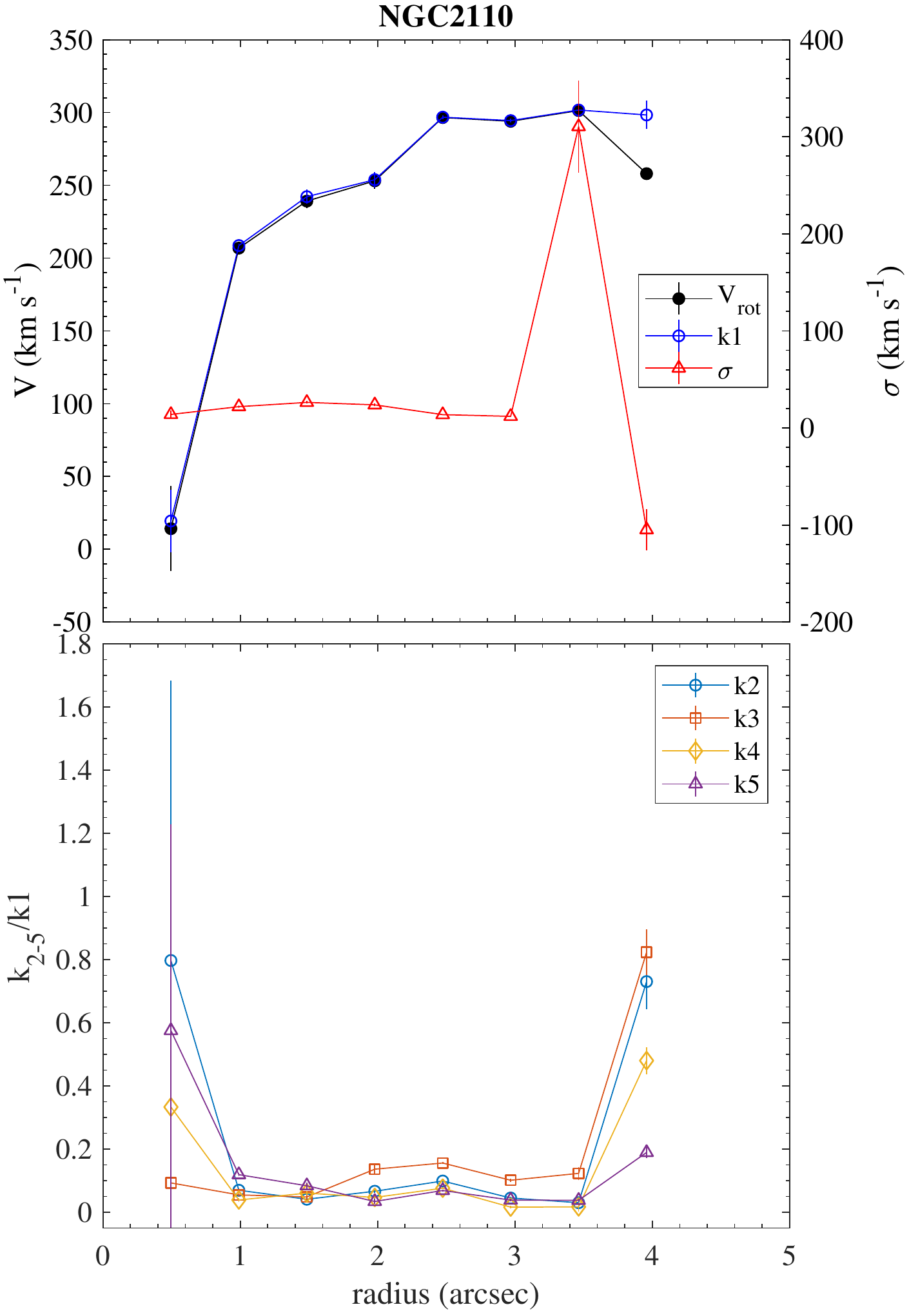}
	\end{subfigure}
	\caption{The same as in Figure~\ref{NGC1386kin2} for NGC~2110.}
	\label{NGC2110kin2}
\end{figure*}

\begin{figure*}
	\centering
	\begin{subfigure}{0.45\textwidth}
	\includegraphics[width=\linewidth]{images/ESO428-G14-velMap.pdf}
	\end{subfigure}
	\begin{subfigure}{0.45\textwidth}
	\includegraphics[width=0.85\linewidth]{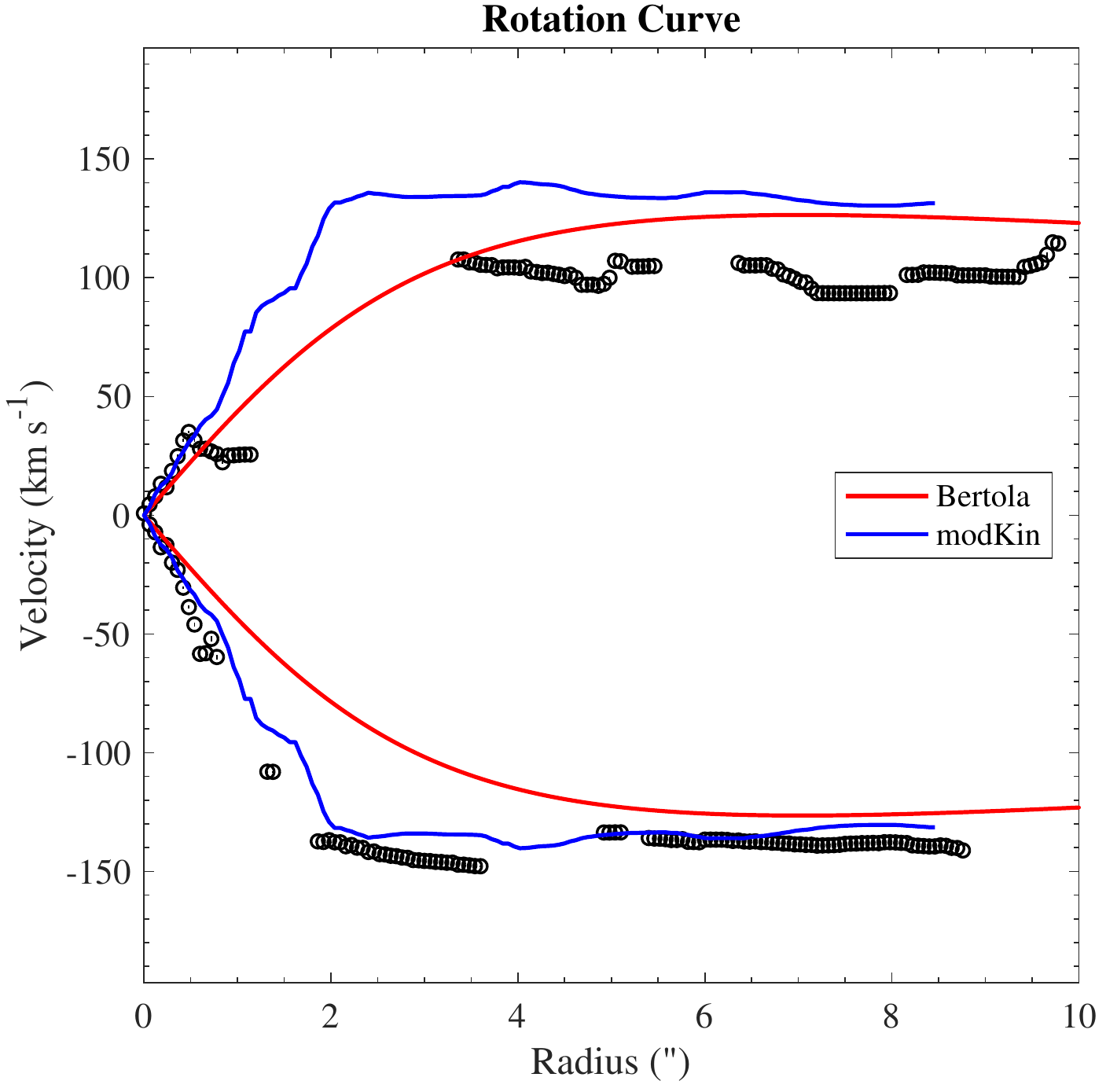}
	\end{subfigure}
	\begin{subfigure}{0.45\textwidth}
	\includegraphics[width=\linewidth]{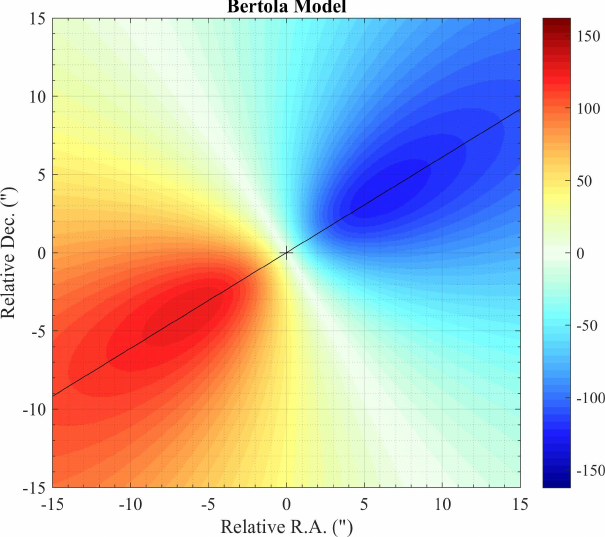}
	\end{subfigure}
	\begin{subfigure}{0.45\textwidth}
	\includegraphics[width=\linewidth]{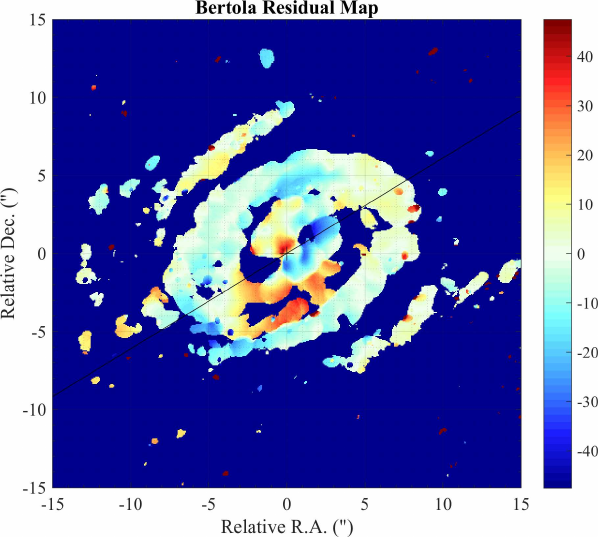}
	\end{subfigure}
	\begin{subfigure}{0.45\textwidth}
	\includegraphics[width=\linewidth]{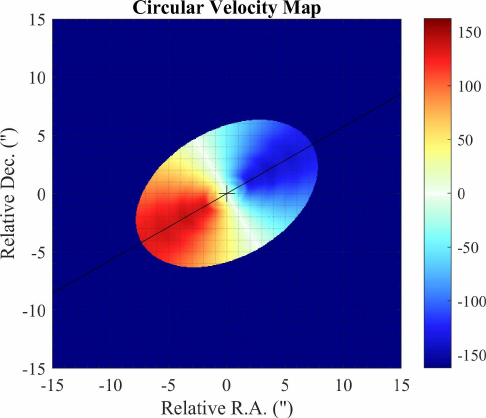}
	\end{subfigure}
	\begin{subfigure}{0.45\textwidth}
	\includegraphics[width=\linewidth]{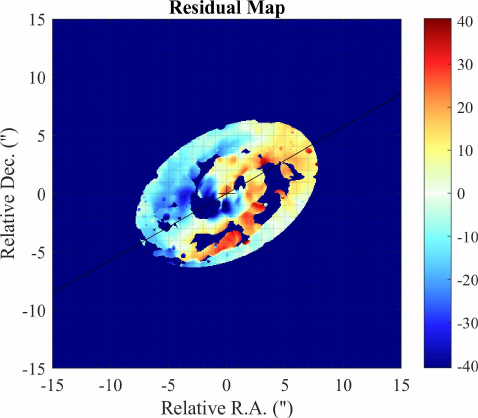}
	\end{subfigure}
	\caption{The same as in Figure~\ref{NGC1386kin1} for ESO~428-G14.}
	\label{ESO428kin1}
\end{figure*}

\begin{figure*}
	\centering
	\begin{subfigure}[b]{0.48\textwidth}
		\includegraphics[width=\columnwidth]{images/ESO428-G14-velMap.pdf}
		\includegraphics[width=\columnwidth]{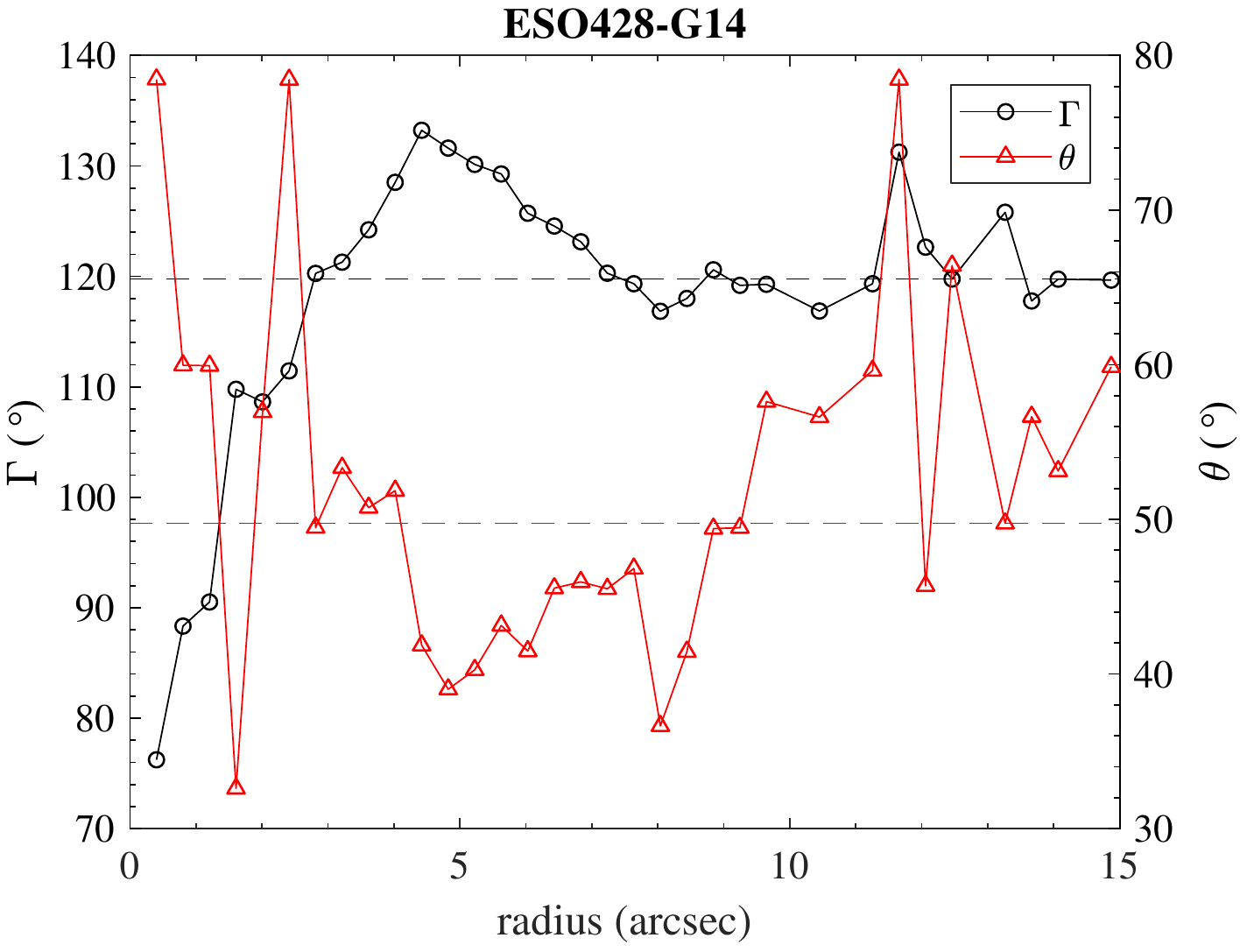}
	\end{subfigure}
	\begin{subfigure}[t]{0.48\textwidth}
		\includegraphics[width=\columnwidth]{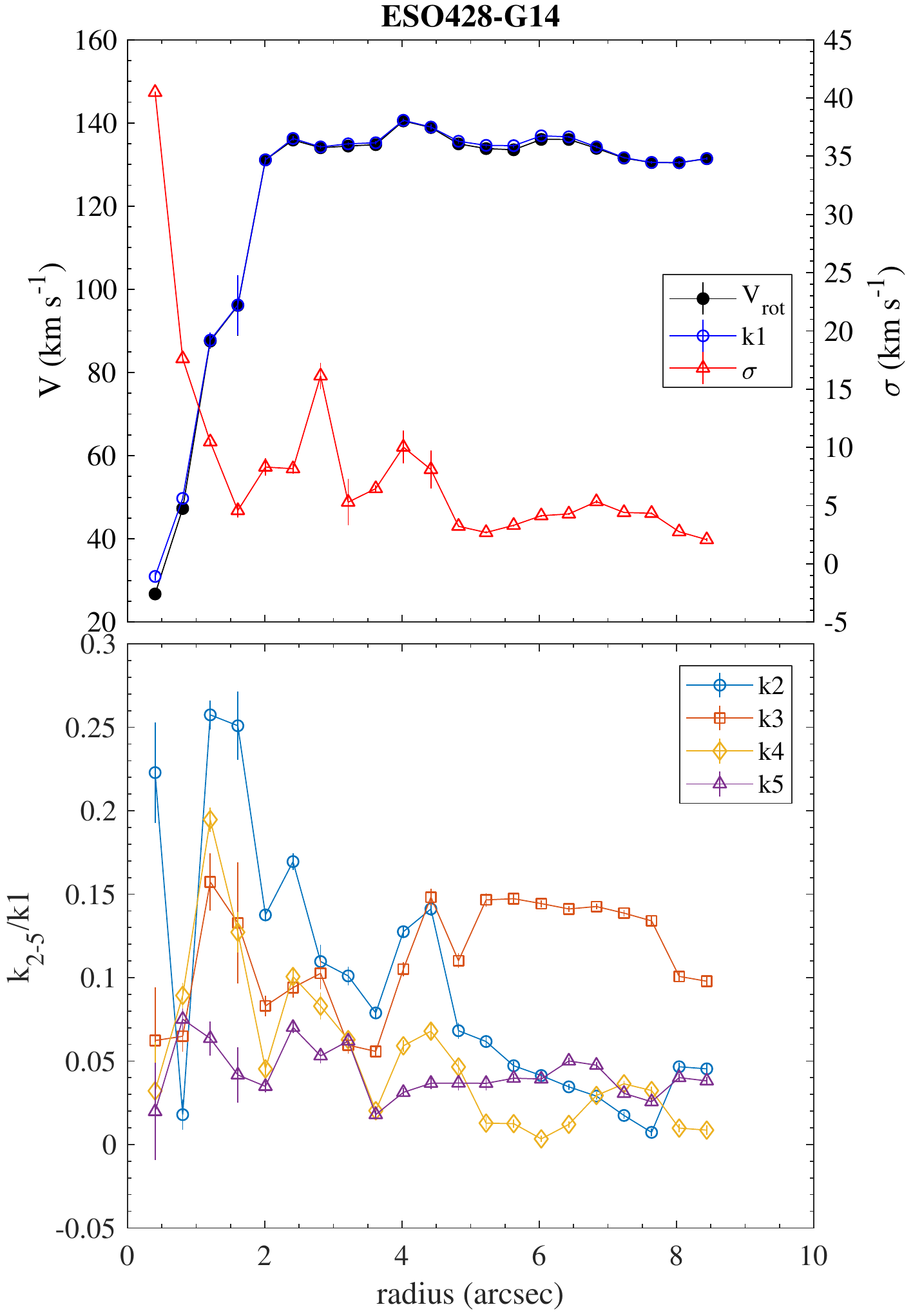}
	\end{subfigure}
	\caption{The same as in Figure~\ref{NGC1386kin2} for ESO~428-G14.}
	\label{ESO428kin2}
\end{figure*}

\begin{figure*}
	\centering
	\begin{subfigure}{0.45\textwidth}
	\includegraphics[width=\linewidth]{images/NGC3081-velMap.pdf}
	\end{subfigure}
	\begin{subfigure}{0.45\textwidth}
	\includegraphics[width=0.85\linewidth]{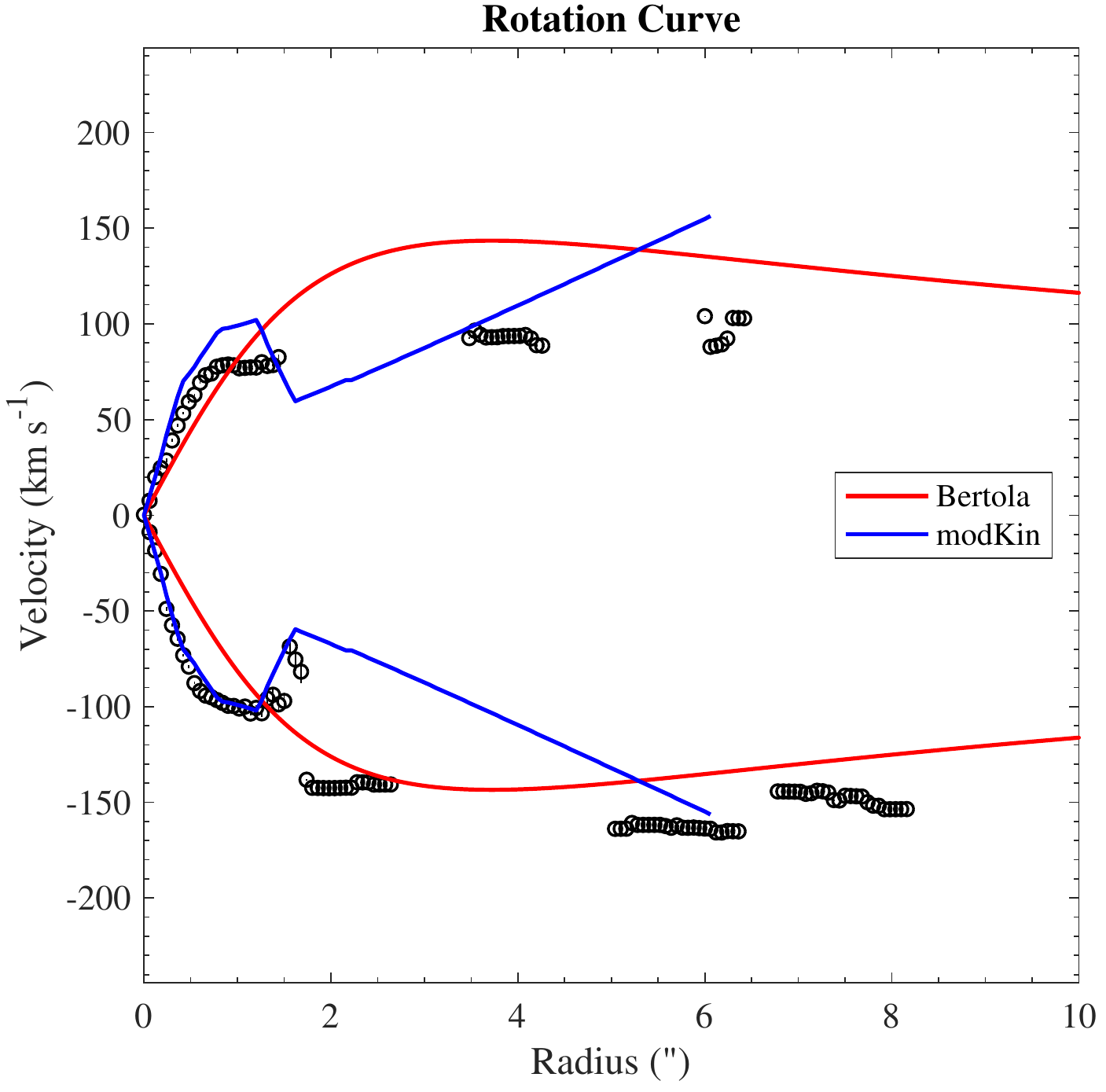}
	\end{subfigure}
	\begin{subfigure}{0.45\textwidth}
	\includegraphics[width=\linewidth]{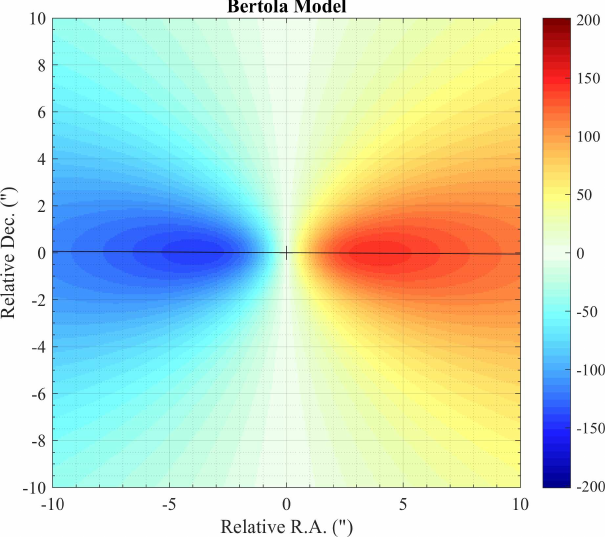}
	\end{subfigure}
	\begin{subfigure}{0.45\textwidth}
	\includegraphics[width=\linewidth]{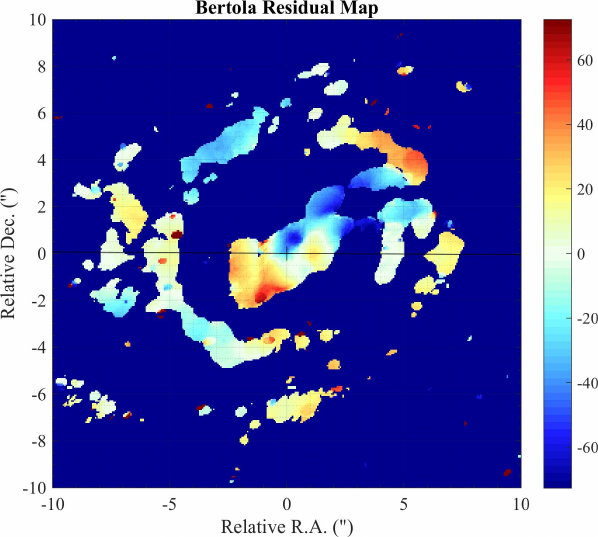}
	\end{subfigure}
	\begin{subfigure}{0.45\textwidth}
	\includegraphics[width=\linewidth]{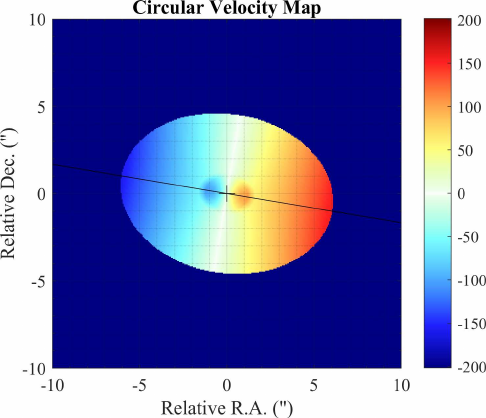}
	\end{subfigure}
	\begin{subfigure}{0.45\textwidth}
	\includegraphics[width=\linewidth]{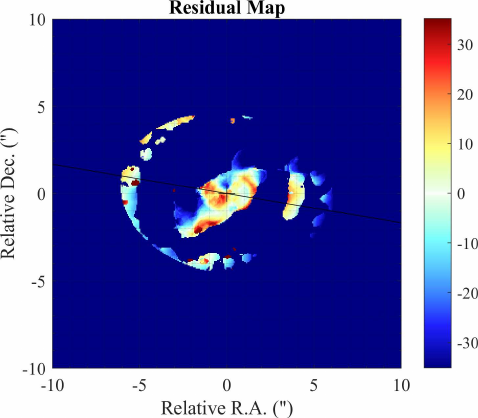}
	\end{subfigure}
	\caption{The same as in Figure~\ref{NGC1386kin1} for NGC~3081.}
	\label{NGC3081kin1}
\end{figure*}

\begin{figure*}
	\centering
	\begin{subfigure}[b]{0.48\textwidth}
		\includegraphics[width=\columnwidth]{images/NGC3081-velMap.pdf}
		\includegraphics[width=\columnwidth]{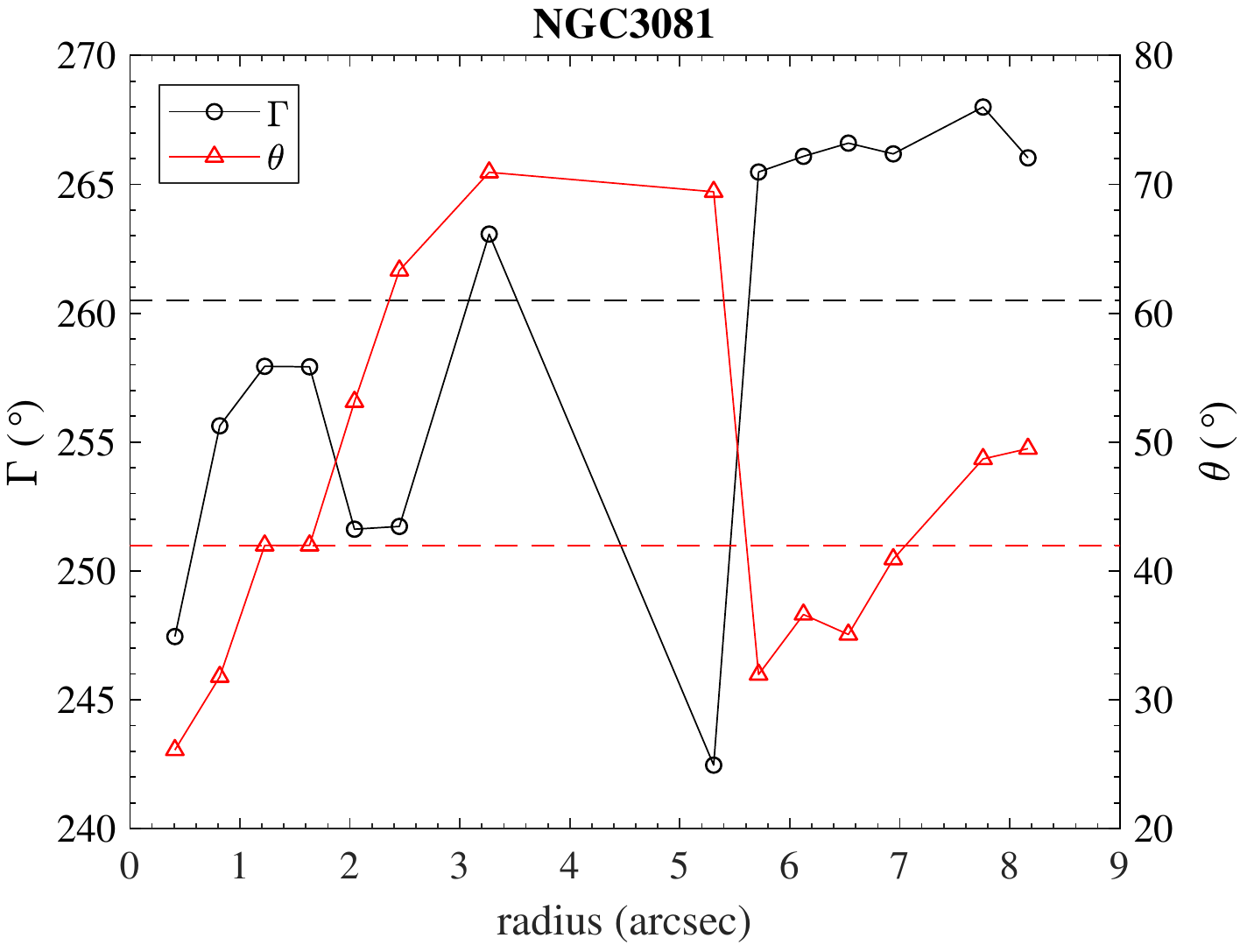}
	\end{subfigure}
	\begin{subfigure}[t]{0.48\textwidth}
		\includegraphics[width=\columnwidth]{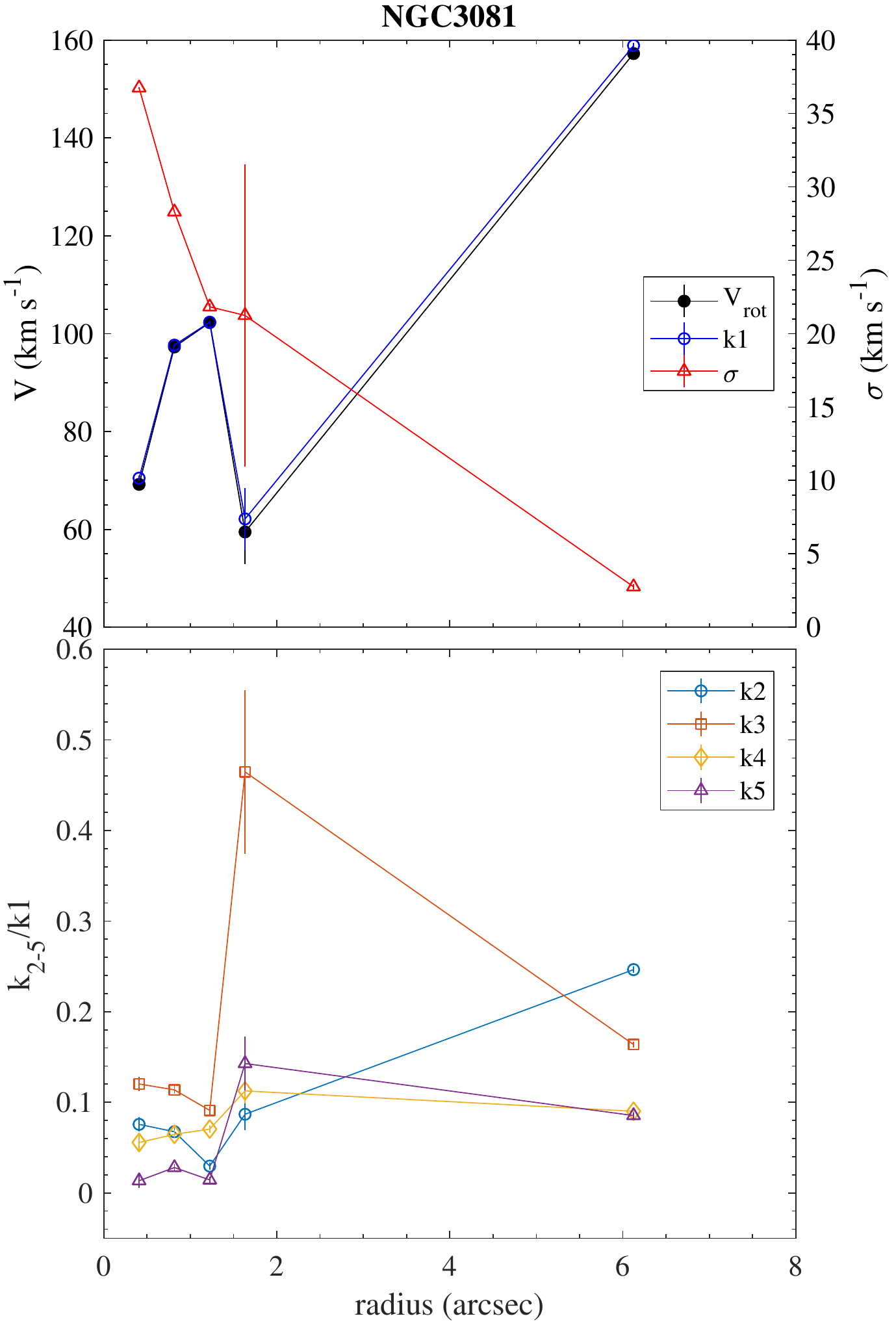}
	\end{subfigure}
	\caption{The same as in Figure~\ref{NGC1386kin2} for NGC~3081.}
	\label{NGC3081kin2}
\end{figure*}

\begin{figure*}
	\centering
	\begin{subfigure}{0.45\textwidth}
	\includegraphics[width=\linewidth]{images/NGC5728-velMap.pdf}
	\end{subfigure}
	\begin{subfigure}{0.45\textwidth}
	\includegraphics[width=0.85\linewidth]{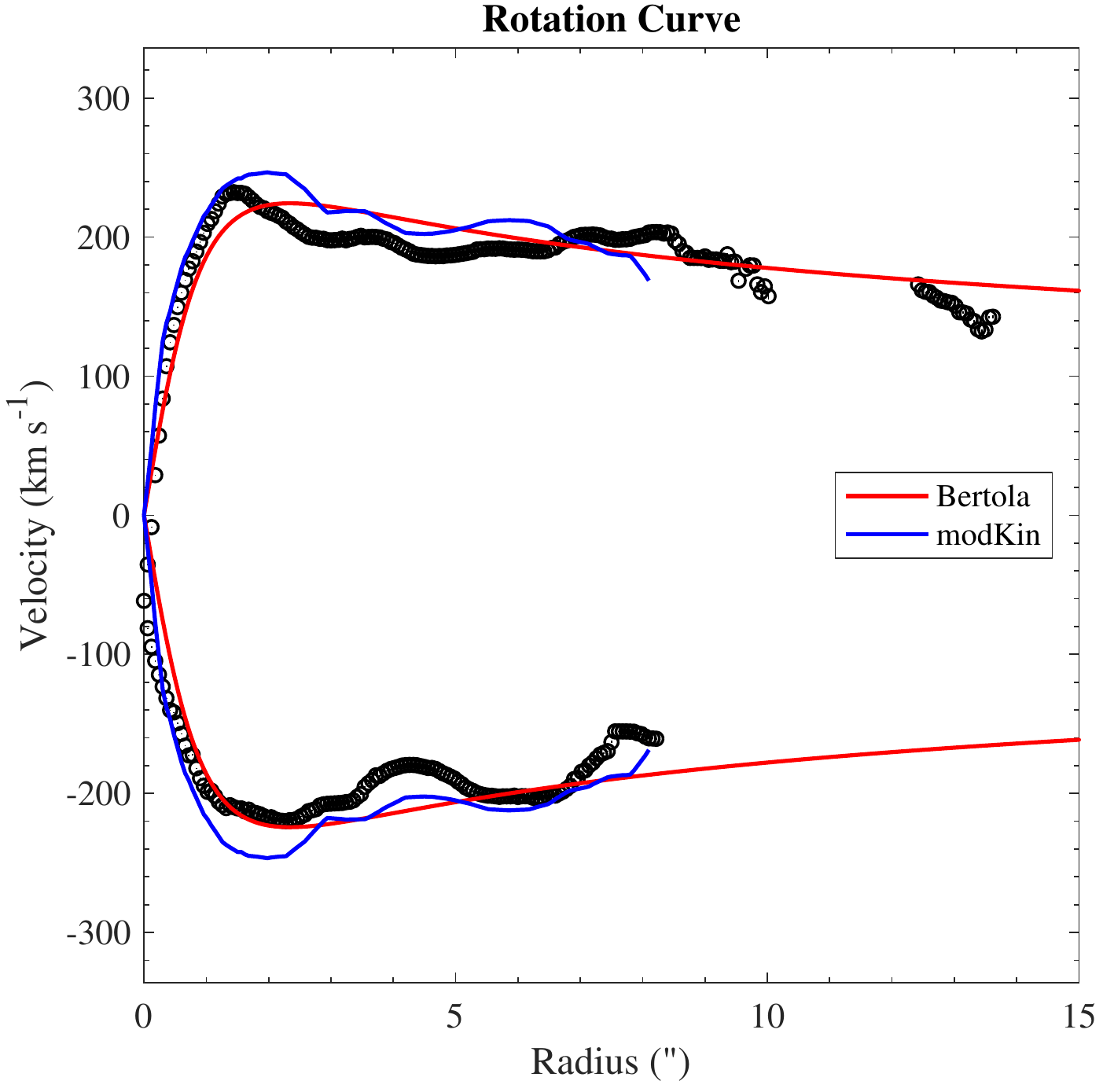}
	\end{subfigure}
	\begin{subfigure}{0.45\textwidth}
	\includegraphics[width=\linewidth]{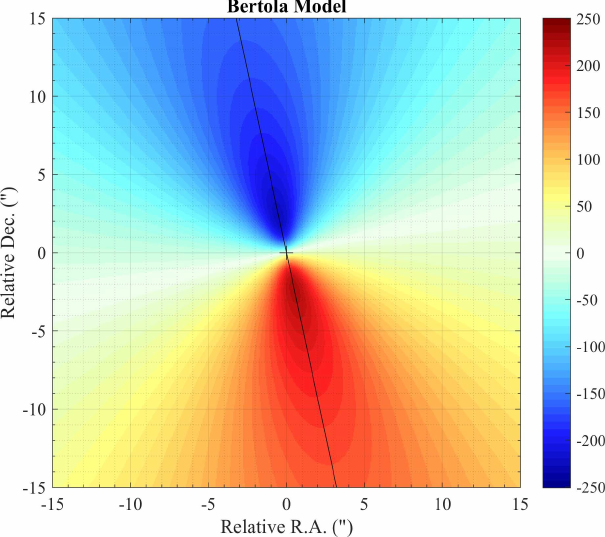}
	\end{subfigure}
	\begin{subfigure}{0.45\textwidth}
	\includegraphics[width=\linewidth]{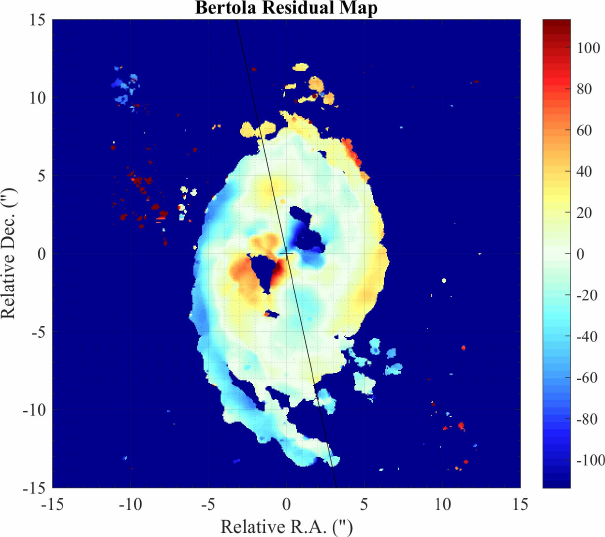}
	\end{subfigure}
	\begin{subfigure}{0.45\textwidth}
	\includegraphics[width=\linewidth]{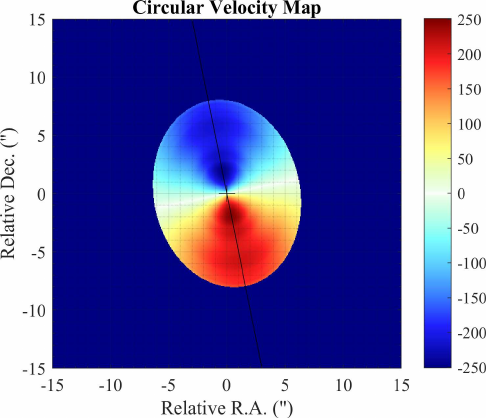}
	\end{subfigure}
	\begin{subfigure}{0.45\textwidth}
	\includegraphics[width=\linewidth]{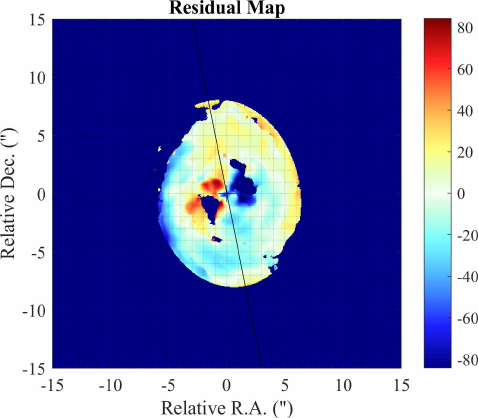}
	\end{subfigure}
	\caption{The same as in Figure~\ref{NGC1386kin1} for NGC~5728.}
	\label{NGC5728kin1}
\end{figure*}

\begin{figure*}
	\centering
	\begin{subfigure}[b]{0.48\textwidth}
		\includegraphics[width=\columnwidth]{images/NGC5728-velMap.pdf}
		\includegraphics[width=\columnwidth]{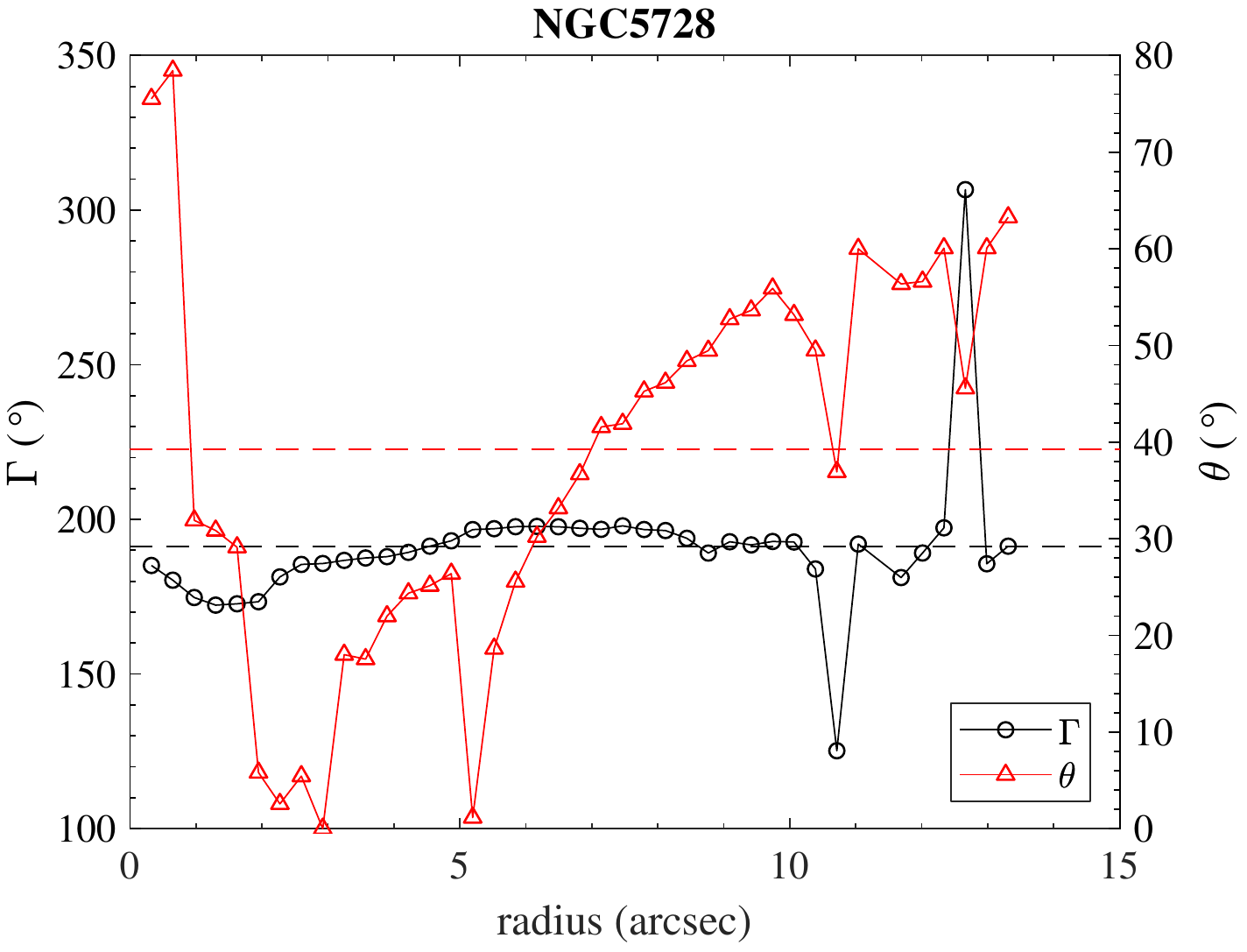}
	\end{subfigure}
	\begin{subfigure}[t]{0.48\textwidth}
		\includegraphics[width=\columnwidth]{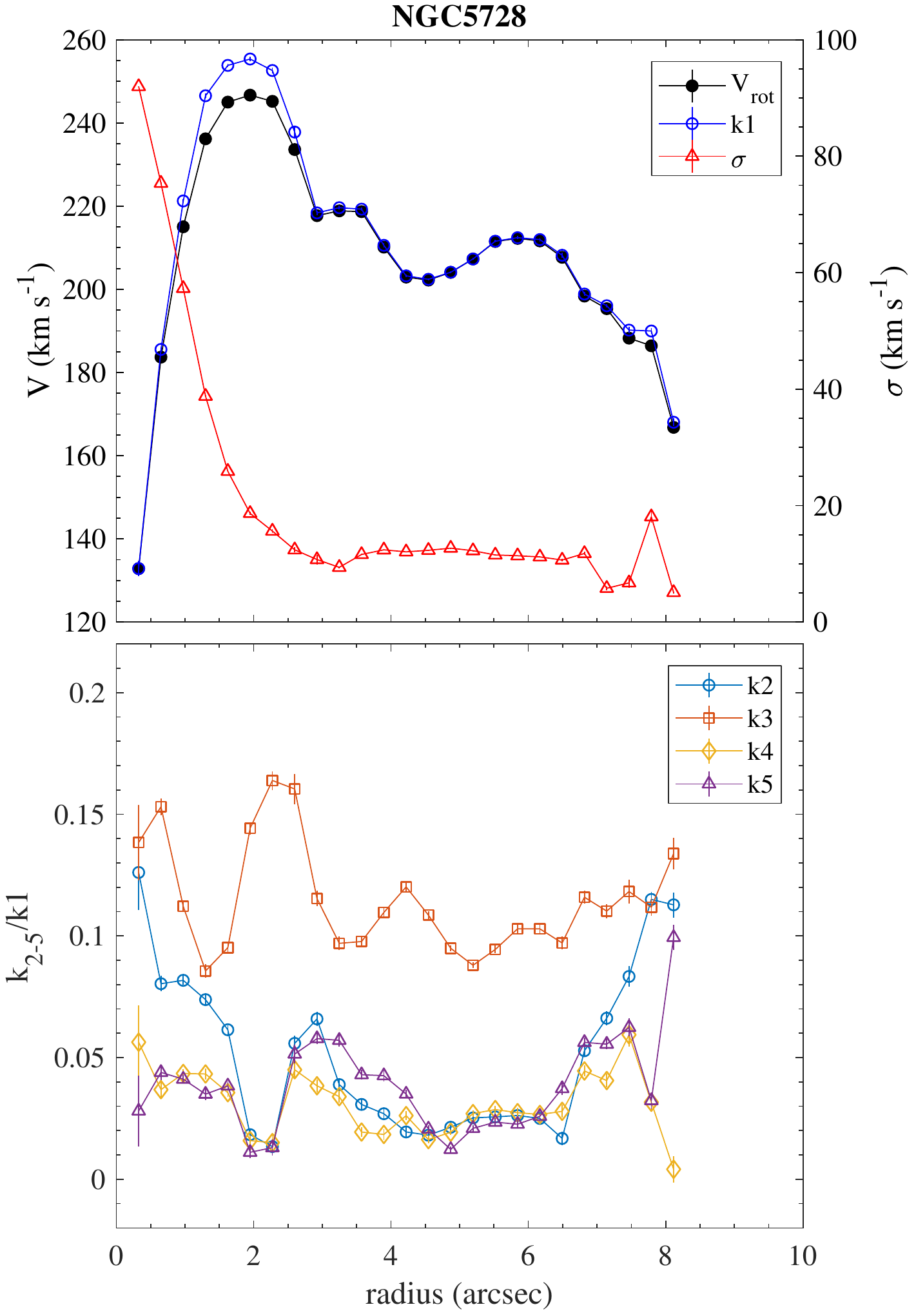}
	\end{subfigure}
	\caption{The same as in Figure~\ref{NGC1386kin2} for NGC~5728.}
	\label{NGC5728kin2}
\end{figure*}

\begin{figure*}
	\centering
	\begin{subfigure}{0.45\textwidth}
	\includegraphics[width=\linewidth]{images/NGC7213-velMap.pdf}
	\end{subfigure}
	\begin{subfigure}{0.45\textwidth}
	\includegraphics[width=0.85\linewidth]{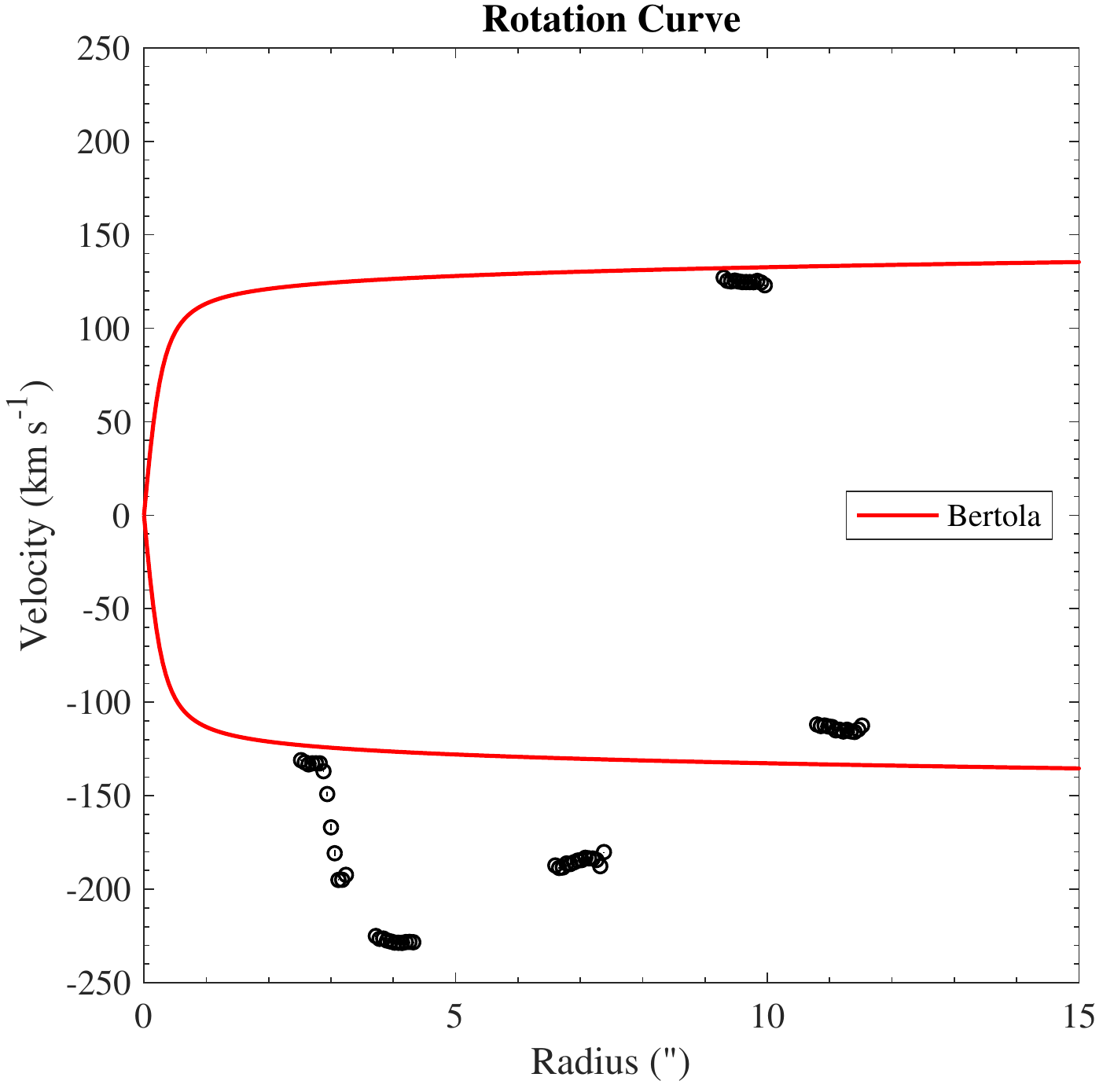}
	\end{subfigure}
	\begin{subfigure}{0.45\textwidth}
	\includegraphics[width=\linewidth]{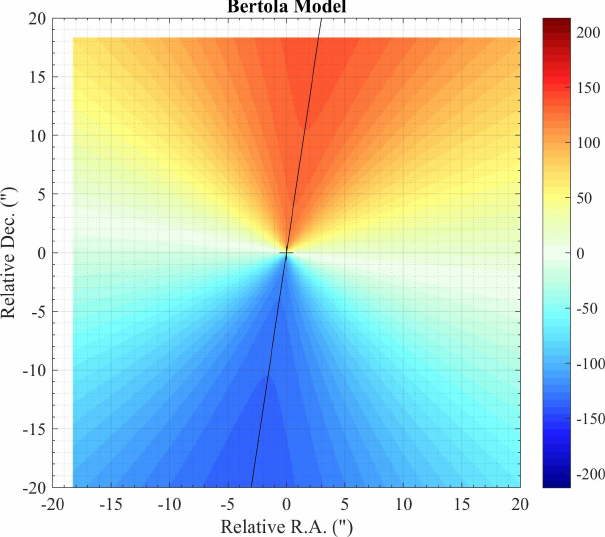}
	\end{subfigure}
	\begin{subfigure}{0.45\textwidth}
	\includegraphics[width=\linewidth]{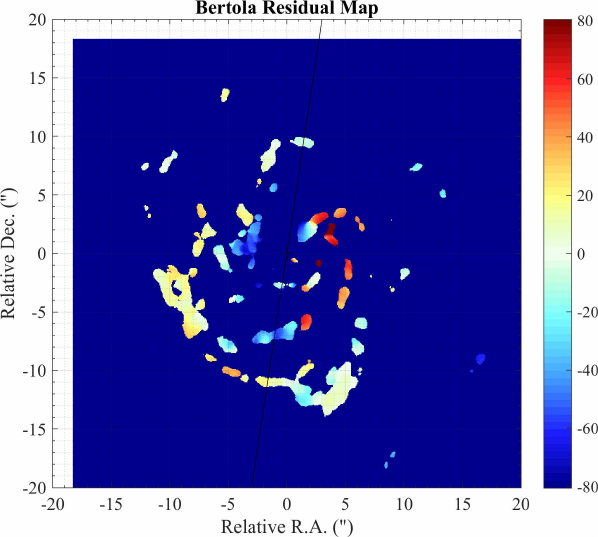}
	\end{subfigure}
	\caption{The same as in Figure~\ref{NGC1386kin1} for NGC~7213 with the exception of model and residual maps from \textit{modKin} analysis.}
	\label{NGC7213kin1}
\end{figure*}

\end{document}